\documentclass[a4paper,11pt]{article}
\pdfoutput=1 
\usepackage{jheparxiv}
\usepackage[T1]{fontenc}
\usepackage[normalem]{ulem}

\usepackage{tensor}
\usepackage{amsmath}
\usepackage{amssymb}
\usepackage{mathrsfs}
\usepackage{graphicx}
\usepackage{stmaryrd}
\usepackage{twistor}
\usepackage{tikz}
\usepackage{commath}
\usepackage{xcolor}
\usepackage{mathtools}
\usepackage{tkz-euclide,subfigure}
\usepackage{capt-of}

\newcommand{\rg}{\mathrm{g}}
\newcommand{\cross}[1]{
    \draw (#1) ++ (-0.1,0.1) -- ++(0.2,-0.2);
    \draw (#1) ++ (-0.1,-0.1) -- ++(0.2,0.2);
}

\renewcommand{\i}{{\rm i}}

\title{Eikonal amplitudes on the celestial sphere}

\author[a]{Tim Adamo,} 
\author[a]{Wei Bu,}
\author[b]{Piotr Tourkine}
\author[a]{\& Bin Zhu}

\affiliation[a]{School of Mathematics and Maxwell Institute for Mathematical Sciences \\
        University of Edinburgh, EH9 3FD, United Kingdom}
\affiliation[b]{LAPTh, CNRS et Universit\'{e} Savoie Mont-Blanc \\
9 Chemin de Bellevue, F-74941 Annecy, France}

\emailAdd{t.adamo@ed.ac.uk}
\emailAdd{w.bu@sms.ed.ac.uk}
\emailAdd{tourkine@lapth.cnrs.fr}
\emailAdd{bzhu@exseed.ed.ac.uk}

\abstract{Celestial scattering amplitudes for massless particles are Mellin transforms of momentum-space scattering amplitudes with respect to the energies of the external particles, and behave as conformal correlators on the celestial sphere. However, there are few explicit cases of well-defined celestial amplitudes, particularly for gravitational theories: the mixing between low- and high-energy scales induced by the Mellin transform generically yields divergent integrals. In this paper, we argue that the most natural object to consider is the gravitational amplitude dressed by an oscillating phase arising from semi-classical effects known as eikonal exponentiation. This leads to gravitational celestial amplitudes which are analytic, apart from a set of poles at integer negative conformal dimensions, whose degree and residues we characterize. We also study the large conformal dimension limits, and provide an asymptotic series representation for these celestial eikonal amplitudes. Our investigation covers two different frameworks, related by eikonal exponentiation: $2\to2$ scattering of scalars in flat spacetime and $1\to1$ scattering of a probe scalar particle in a curved, stationary spacetime. These provide data which any putative celestial dual for Minkowski, shockwave or black hole spacetimes must reproduce. We also derive dispersion and monodromy relations for these celestial amplitudes and discuss Carrollian eikonal-probe amplitudes in curved spacetimes.}

\begin{document}
\maketitle
\flushbottom

\section{Introduction}

When computing scattering amplitudes in quantum field theory (QFT), the external states of a given scattering process are usually expressed in a momentum eigenstate basis. While these momentum eigenstates manifest many of the intuitive physical features of scattering amplitudes and are naturally compatible with the Wilsonian paradigm of QFT in terms of energy scales, one can in fact represent the external states using {any} basis of solutions to the free equations of motion. A particularly interesting alternative are \emph{conformal primary states}, which, in four spacetime dimensions, are parametrized by a conformal dimension and a point on the celestial two-sphere of null directions and transform as two-dimensional conformal primaries on the celestial sphere under the action of the Lorentz group~\cite{Pasterski:2016qvg,Pasterski:2017kqt,Pasterski:2017ylz}. For massless states, these conformal primary wavefunctions are related to momentum eigenstates by a Mellin transform with respect to the energy. On purely kinematic grounds, a scattering amplitude evaluated with conformal primary external states is called a \emph{celestial amplitude} and will behave like a conformal correlator on the celestial sphere.

However, as pointed out in~\cite{Arkani-Hamed:2020gyp}, 
and despite being the subject of intense study in recent years, most examples of celestial amplitudes studied in the literature always present either some divergent behaviour, or at least some form of severe non-analyticity. This comes from the fact that the Mellin transform which defines any massless celestial amplitude sweeps out the entire energy scale of the external states, mixing the infrared (IR) and ultraviolet (UV) regimes~\cite{Arkani-Hamed:2020gyp}. Consequently, the Mellin transform of a QFT momentum space scattering amplitude evaluated at some finite order in perturbation theory generically leads to divergent integrals and celestial amplitudes which are at worst un-defined and at best distributional~\cite{Pano:2024eek}\footnote{Indeed, \cite{Pano:2024eek} observed that celestial amplitudes can always be defined as distributions when paired with suitable wavepackets, although the nature of this function space depends on the quantum numbers of the external states.}. For example, the scattering of massless scalars, minimally coupled to gravity, produces scattering amplitudes with divergent UV-behaviour at any order in perturbation theory, and the corresponding celestial amplitudes are un-defined.

String theory provides an important framework where analytic, or rather meromorphic, celestial amplitudes can be obtained~\cite{Stieberger:2018edy,Chang:2021wvv,Donnay:2023kvm,Castiblanco:2024hnq} thanks to UV softness, but the lack of explicit field theory examples is surprising. Some analytic properties of celestial amplitudes in complete theories of quantum gravity have been deduced based on general considerations (e.g., that high-energy scattering should be dominated by black hole production) \cite{Stieberger:2018edy,Arkani-Hamed:2020gyp,Donnay:2023kvm}, but in the absence of explicit examples beyond string theory.

\medskip

In this paper, we argue that rather than considering amplitudes to some fixed order in perturbation theory, it is more natural to consider amplitudes to {all-orders} in a particular regime where this is tractable: namely, the eikonal approximation (see \cite{Levy:1969cr,DiVecchia:2023frv} for reviews). We focus on the case of gravitational scattering of massless scalars, which is UV-divergent at any fixed order in Newton's constant, $G$. In momentum space, the eikonal approximation corresponds to a classical iteration of the tree-level approximation which can be re-summed as an exponential series. This is known to regulate the UV behaviour of the amplitude, and we demonstrate that this allows for proper definitions of the Mellin transform mixing the UV and IR regimes.

\medskip

Before describing this new approach, let us first discuss existing strategies for the study of celestial amplitudes. Given the lack of well-defined field theory examples at finite orders in perturbation theory, one can ask how it has been possible to say anything about them before now. There have been essentially three routes towards avoiding this underlying problem. The first is to consider universal features of a celestial amplitude which \emph{are} generically well-defined. For instance, celestial amplitudes exhibit universal factorization properties when the conformal dimensions of external gluons or gravitons approach special values~\cite{Donnay:2018neh,Fan:2019emx,Pate:2019mfs, Adamo:2019ipt,Puhm:2019zbl,Guevara:2019ypd}, or when the insertion locations of two external states on the celestial sphere approach one another~\cite{Fan:2019emx,Pate:2019lpp,Banerjee:2020kaa,Fotopoulos:2019vac,Fotopoulos:2020bqj,Banerjee:2020zlg,Banerjee:2020vnt,Adamo:2021zpw}. The resulting `conformal soft theorems' and `celestial operator product expansion coefficients' are the celestial analogies of the usual soft gluon/graviton theorems and collinear splitting functions in momentum space, and are meromorphic functions of the conformal dimensions. The conformal structures manifested by celestial amplitudes have led to the discovery of many new features of massless scattering, such as the existence of infinite-dimensional symmetry algebras associated with the chiral sectors of gauge theory and gravity~\cite{Guevara:2021abz,Strominger:2021mtt,Himwich:2021dau}.

The second strategy is to consider special cases where the celestial amplitude is defined as a distribution. A prominent example of this is for gluon/photon mediated scattering: with all external states taking values in the principal continuous series of SL$(2,\C)$, the celestial amplitudes are distributional in the imaginary parts of the conformal dimensions~\cite{Pasterski:2017ylz,Schreiber:2017jsr}. While this leads to concrete, distributional expressions for the celestial amplitudes, they are non-analytic in the conformal dimensions, with a delta-function enforcing 4-dimensional dilatation invariance.

The third strategy is to regularize celestial amplitudes by switching on a background field~\cite{Fan:2022vbz,Casali:2022fro,Fan:2022kpp,deGioia:2022fcn,Gonzo:2022tjm,Banerjee:2023rni,Ball:2023ukj,Crawley:2023brz}. This has led to many interesting formulae with close connections to limits of Liouville theory~\cite{Stieberger:2022zyk,Taylor:2023bzj,Stieberger:2023fju,Giribet:2024vnk} and AdS correlators~\cite{Casali:2022fro,deGioia:2022fcn}, but the backgrounds are typically scalar in nature and the fundamental reason for their appearance remains unclear. 

\medskip
In this paper, we propose a new way to generate well-defined celestial amplitudes in field theory via the eikonal approximation. In momentum space, this corresponds to the limit of small angle (or large impact parameter) scattering at high energies. More precisely, if $E$ is the energy scale of the incoming particles and $q$ is the momentum transfer between them, then the eikonal approximation for gravitational scattering is characterised the the hierarchy of scales (cf., \cite{DiVecchia:2023frv}): $\frac{\hbar}{E}\ll G\, E \ll\frac{\hbar}{q}$, where $\hbar$ is the reduced Planck's constant and $G$ is Newton's constant. This hierarchy implies both the small-angle approximation ($q\ll E$)\footnote{Or, equivalently, the large impact parameter approximation ($b\sim \hbar/q$).} and the `trans-Planckian energy' condition $G\,E^2\gg\hbar$, with the latter essentially enforcing the classical limit.

In the eikonal approximation, the perturbative series can be resummed as an exponential, leading to a scattering amplitude which is equal to the tree-level, Born approximation dressed by a phase. This eikonal exponentiation holds both for $2\to2$ gravitational scattering of massless scalars in Minkowski spacetime as well as for $1\to1$ massless probe scattering in any stationary spacetime (or its ultrarelativistic limit), and it has long been known that the two are closely related~\cite{tHooft:1987vrq,Amati:1987uf,Kabat:1992tb,Adamo:2021rfq}. The resulting momentum space amplitudes, which we refer to as \emph{eikonal} (for $2\to2$ in flat space) and \emph{eikonal-probe} (for $1\to1$ in curved space) are all-orders in the gravitational coupling, but essentially perturbative in nature, being the resummation of iterated Born approximations.

Thanks to the oscillating eikonal phase, the Mellin transforms of these amplitudes with respect to the energies of the external scalars have several remarkable properties. They are meromorphic functions of the conformal dimensions, with an infinite sequence of poles along the negative real axis that we characterize. In addition, we analyze the asymptotic behaviour of these celestial amplitudes in the limit of large conformal dimensions. We also provide compact, asymptotic series representations of the amplitudes in terms of certain shift operators acting on the conformal dimensions.

Furthermore, the dependence on all scales (e.g., coupling or background energies) becomes essentially trivial, and can be factorized out of both the celestial eikonal and eikonal-probe amplitudes. This is a result of the Mellin transform which mixes the UV and IR regimes of the momentum space amplitude and does not generally admit a perturbative limit. To our knowledge, celestial eikonal amplitudes are the first explicit examples of meromorphic, non-perturbative gravitational celestial amplitudes\footnote{We note that explicit celestial amplitudes coupled to gravity have been constructed in the context of 2-dimensional integrable field theories~\cite{Duary:2022onm,Kapec:2022xjw}. Our focus in this work is on four spacetime dimensions.}. The celestial eikonal-probe amplitudes, which we consider on gravitational shockwave, Schwarzschild and Kerr spacetimes, provide crucial data which any putative celestial CFT dual for these spacetimes must reproduce. There is also a precise correspondence between the celestial eikonal amplitude and the celestial eikonal-probe amplitude on a shockwave background, following from the close connection between the two in momentum space~\cite{tHooft:1987vrq}. 

By analytically continuing the underlying Mellin transforms, we obtain dispersion and monodromy relations for the celestial eikonal and eikonal-probe amplitudes, respectively. We show that eikonal exponentiation also resolves divergences appearing in Carrollian amplitudes, which are position-space amplitudes defined at the conformal boundary of an asymptotically flat spacetime and related to celestial amplitudes by a certain modified Mellin transform~\cite{Bagchi:2022emh,Donnay:2022aba,Donnay:2022wvx,Mason:2023mti,Salzer:2023jqv,Nguyen:2023miw,Bagchi:2023cen,Liu:2024nfc,Have:2024dff,Stieberger:2024shv}.

\medskip

In hindsight, the key role played by the phase of eikonal exponentiation should be no surprise. As shown in a series of papers by Amati-Ciafaloni-Veneziano (ACV)~\cite{Amati:1987uf,Amati:1987wq,Amati:1988tn,Amati:1990xe}\footnote{See also the recent review~\cite{DiVecchia:2023frv}.}, eikonal exponentiation is a general feature of the gravitational S-matrix. Choosing to work at some fixed order in perturbation theory is un-natural from the perspective of celestial amplitudes, where energy scales are mixed together by the Mellin transform. The eikonal approximation captures the first appearance of universal, all-orders information in the full gravitational scattering amplitude. This suggests that when treated properly, celestial gravitational scattering amplitudes \emph{are} naturally meromorphic functions.

Let us comment on two important points about our use of the eikonal exponentiation. Eikonal exponentiation is usually thought of as emerging in the strict small-angle and trans-Planckian energy limits. Yet to perform the Mellin transform to obtain celestial amplitudes, we typically must work at some \emph{fixed} angle, which can be small but is non-zero. Furthermore, the Mellin transform runs over the full energy scale, including small (i.e., non-trans-Planckian) energies which lie outside the strictly classical regime of eikonal scattering. One could then ask to what extent such amplitudes make sense on the celestial sphere. 

In the case of small, but fixed angle scattering, this question was studied by ACV in~\cite{Amati:1987uf,Amati:1987wq,Amati:1988tn,Amati:1990xe}. There, it was shown that, due to the softness of the gravitational interaction, the regime of validity for the eikonal amplitude can be extrapolated from vanishing to small, but finite scattering angles. This can be seen to stem from the fact that the gravitationnal interaction reduces to a series of successive small re-scatterings, which accumulate to scattering at finite angle. Indeed, a single scattering produces an infinitesimal, quantum deflection angle of order $\hbar/(b\,E)$, where $b$ is the impact parameter, while the eikonal resummation leads to a finite classical deflection angle of order $GE/b$. This fact is also closely related to the correspondence between eikonal $2\to2$ scattering and $1\to1$ probe scattering on a shockwave background, which exhibits eikonal exponentiation for finite scattering angles~\cite{tHooft:1987vrq}. 

As for integrating the eikonal amplitude over the full energy scale, the regime of validity of the eikonal approximation means that any features of the resulting celestial amplitude which depend on the small energy (i.e., $GE^2\lesssim\hbar$) region of integration will generically receive quantum corrections\footnote{We thank Carlo Heissenberg for emphasizing this point to us and for discussions on extension to finite scattering angles.}. In particular, this will be true of the pole structure of the amplitude, which we show arises from precisely this low-energy region. However, the key role of the exponentiation (as opposed to the Born approximation) is to regulate the \emph{high} energy region of the Mellin transform, which is precisely where the classical constraint of the eikonal approximation holds. Furthermore, in the limit of large conformal dimensions, the Mellin transform is dominated by a saddle at high energy, which we can evaluate explicitly to obtain a result consistent with the classical limit. 

\medskip

Finally, some important aspects of celestial eikonal or eikonal-probe amplitudes have been considered before in the literature. In~\cite{deGioia:2022fcn}, the eikonal approximation for $2\to2$ scattering in the conformal primary basis was characterized by taking small conformal cross ratio and large conformal dimensions, although the convergence of the Mellin integrals and analytic properties of celestial eikonal amplitudes were not investigated. This is completely consistent with our findings, which show that the Mellin transform of an eikonal amplitude is dominated by a classical (i.e., trans-Planckian energy) saddle for large conformal dimensions. The authors of~\cite{deGioia:2022fcn} also suggested a correspondence with celestial eikonal-probe scattering on a gravitational shockwave by considering leading terms in the perturbative expansion on both sides. 

The leading perturbative contribution to celestial eikonal-probe amplitudes was considered in~\cite{Gonzo:2022tjm}, where the resulting Mellin amplitudes were observed to be divergent in every case other than Kerr, where the spin of the black hole provides a regulator. Celestial eikonal-probe amplitudes have also been interpreted in terms of three-point correlation functions between conformal primary states and operator insertions on the celestial sphere encoding the background~\cite{Pasterski:2020pdk,Gonzo:2022tjm}, although all of these studies have been perturbative in nature. There has also been a study of two-dimensional celestial eikonal amplitudes arising from near-horizon scattering in a black hole background~\cite{Fernandes:2023ibv}.

\medskip

The paper is organized as follows. In Section~\ref{sec:4pt_celestial}, we introduce celestial amplitudes corresponding to 4-point flat-space S-matrices. We then introduce eikonal amplitudes and study their Mellin transform, proving that the celestial eikonal amplitude is well-defined as a meromorphic function of the conformal dimensions and studying its analytic properties. Section~\ref{sec:2pt_celestial} considers a similar story for 2-point eikonal-probe amplitudes in asymptotically flat stationary spacetimes, establishing the analytic properties of their Mellin transforms as well as a correspondence with 4-point celestial amplitudes.

In Section~\ref{sec:dispersion} we derive dispersion and monodromy relations for celestial eikonal and eikonal-probe amplitudes, respectively. Section~\ref{sec:carrollian_probe} discusses the role of eikonal exponentiation in regularising Carrollian probe amplitudes on curved, asymptotically flat spacetimes. Section~\ref{sec:discussion} comments briefly on going beyond the leading eikonal approximation due to subleading eikonal resummations and stringy effects, and discusses several directions for future research. Appendix~\ref{appendix:eik_gamma} includes details of the analytic properties of the prototype Mellin integral that we encounter throughout this paper.


\section{Celestial eikonal amplitudes}\label{sec:4pt_celestial}

Typically, the external states of scattering amplitudes are expressed in terms of a momentum eigenstate basis. However, these external states can be parametrized in terms of {any} basis of solutions to the free equations of motion. The \emph{conformal primary basis} is a particularly interesting alternative, which renders -- on purely kinematic grounds -- the resulting scattering amplitude in a form which transforms as a conformal correlator on the celestial sphere. As such, scattering amplitudes expressed in the conformal primary basis are usually referred to as \emph{celestial amplitudes}.

For massless external states, the relationship between momentum space and celestial amplitudes is particularly simple. Parametrizing a null 4-momentum in terms of a frequency $\omega$ and a point\footnote{Throughout, we work with complex stereographic coordinates on the celestial sphere, with $z=\e^{\im\phi}\,\cot(\theta/2)$.} on the celestial sphere $(z,\bar{z})$; in Cartesian coordinates,
\be\label{csnullmom}
p_{\mu}=\left(p_0,p_x,p_y,p_z\right)=\frac{\omega}{\sqrt{2}}\left(1+|z|^2,\,z+\bar{z},\,-\im(z-\bar{z}),\,1-|z|^2\right)\,.
\ee
The $n$-point celestial amplitude is then defined by taking the Mellin transform of the $n$-point momentum space amplitude with respect to the frequencies of the external particles:
\be\label{campdef}
\widetilde{\cM}_n(\{\Delta_i,z_i,\bar{z}_i\})=\int_{0}^{\infty}\prod_{j=1}^{n}\d\omega_i\,\omega_i^{\Delta_i-1}\,\cM_{n}(\{\omega_i,z_i,\bar{z}_i\})\,.
\ee
The celestial amplitude is defined for all $\Delta_i\in\C$, although a complete and normalizable basis of conformal primary states is constituted by the principal continuous series $\Delta_i=1+\im\,\lambda_i$ for $\lambda_i\in\R$. It is straightforward to show that the celestial amplitude $\widetilde{\cM}_{n}$ behaves as a conformal correlator on the celestial sphere, with primaries of conformal dimension $\Delta_i$ inserted at locations $(z_i,\bar{z}_i)$. 

This observation has motivated a growing effort to leverage these conformal structures to learn more about massless scattering amplitudes and, most ambitiously, work towards a holographic formulation of a dual CFT on the sphere which actually computes celestial amplitudes dynamically. However, the fact that the Mellin transform in \eqref{campdef} sweeps out the full energy scale of $\omega\in(0,\infty)$ means that celestial amplitudes mix the ultraviolet (UV) and infrared (IR) physics of the underlying scattering process. This is problematic for theories which are not UV complete but nevertheless of physical interest: a particularly striking example is the $2\to2$ scattering of gravitationally-coupled massless scalars. 

In this section, we recall the well-known problems with gravitational celestial amplitudes due to UV/IR mixing from the Mellin transform, and demonstrate that this can be resolved by considering the \emph{eikonal amplitude}, which is all-orders in Newton's constant, $G$, and dresses the Born approximation to the momentum space amplitude by a phase. This phase renders the Mellin transform well-defined, and the result can be compactly approximated in the form of an asymptotic series of shift operators acting on the conformal dimensions. To our knowledge, this is one of the first examples of a well-defined, all-orders gravitational celestial amplitude in 4-dimensional field theory. We also comment on the analytic structure of the celestial eikonal amplitude and its asymptotic behaviour in the limit of large conformal dimensions.


\subsection{Celestial amplitudes and UV/IR mixing}

Let us consider $2\to2$ scattering of massless scalars in a relativistic QFT, denoting the incoming momenta $p_1$, $p_2$ and outgoing momenta $p_1'\equiv p_3$, $p_2'\equiv p_4$, respectively. The kinematic invariants describing the scattering process in this $2\to2$ channel are the Mandelstam variables
\be\label{mandelstam}
s:=(p_1+p_2)^2\,, \qquad t:=(p_1-p_1')^2:=q^2\,, \qquad q:=p_1-p_1'\,.
\ee
The corresponding massless 4-point scattering amplitude can be expressed in momentum space in terms of the 4-point kinematics and a momentum conserving delta function:
\begin{equation}
    \cM_4 =\delta^4\!\left(\sum_{i=1}^4 p_i\right) \,\mathrm{M}(s,t) \, ,
\end{equation}
where $s,t$ are the Mandelstam variables \eqref{mandelstam}.

Parametrizing the four massless external momenta in terms of $(\omega_i,z_i,\bar{z}_i)$ for $i=1,\ldots,4$ as in \eqref{csnullmom}, it is straightforward to show that the celestial 4-point amplitude in the channel $12\rightarrow34$ can be written as~\cite{Stieberger:2018edy,Gonzalez:2020tpi,Arkani-Hamed:2020gyp}:
\begin{equation}
    \widetilde{\cM}_4(\Delta_i,z_i,\bar{z}_i) =\int_{0}^{\infty}\prod_{i=1}^4\d\omega_i\,\omega_i^{\Delta_i-1}\,\cM_4= X\, \mathcal{A}(\beta, z) \, , \label{eq:4-ptMtilde}
\end{equation}
where 
\begin{equation}
    \beta := \sum_{i=1}^4 (\Delta_i -1) \,, \qquad z:=\frac{z_{13}\,z_{24}}{z_{12}\,z_{34}}\,, \label{eq:defbeta}
\end{equation}
and $X$ is a universal kinematic factor
\be\label{Xfactor}
X:=2^{-\beta-2}\,\delta(z-\bar{z})\,|z(1-z)|^{\frac{\beta+4}{6}}\,\prod_{i<j}|z_{ij}|^{\frac{\beta+4}{3}-\Delta_i-\Delta_j}\,.
\ee
Observe that the delta function setting $z=\bar{z}$ is a kinematic constraint, ensuring a real scattering angle. All of the dynamics of the amplitude is encoded in the quantity $\cA(\beta,z)$, which is defined by 
\begin{equation}\label{AAmp}
    \mathcal{A}(\beta,z) = \int_0^{\infty} \d\omega \, \omega^{\beta-1} \, \mathrm{M}(s=\omega^2,\, t= -z\omega^2) \, ,
\end{equation}
with $\omega$ indicating the collective energy scale of the incoming particles and $t$ the transverse exchanged energy.

It is interesting -- and important -- to explore the extent to which the Mellin transform \eqref{AAmp}, encoding the dynamical content of the celestial amplitude is actually well-defined. The tree-level contribution to the $12\to34$ channel is simply given by the single-exchange Born approximation:
\be\label{JspinBAmp}
\mathrm{M}^J_{\mathrm{Born}}(s,t)=\rg^2\,\frac{s^J}{t}=-\rg^2\,\frac{\omega^{2J-2}}{z}\,,
\ee
where $\rg$ is the coupling constant (including numerical factors) between the massless scalars and the massless spin-$J$ exchanged particle. The corresponding celestial amplitude
\be\label{JspinCAmp}
\cA^J_{\mathrm{Born}}(\beta,z)=-\frac{\rg^2}{z}\int_{0}^{\infty}\d\omega\,\omega^{\beta+2J-3}=-\frac{\rg^2}{z}\int_{-\infty}^{\infty}\d t\,\e^{(\beta+2J-2)\,t}\,,
\ee
which is, for generic values of $\beta$, an un-defined integral which diverges exponentially. The only exception is if $\beta+2J-2=\im\lambda$ for some $\lambda\in\R$, in which case $\cA_{\mathrm{Born}}\propto\delta(\lambda)$ and the celestial amplitude is defined as a distribution\footnote{An example of this situation which has a physical interpretation is provided by charged scalars scattering via photon exchange ($J=1$) with conformal dimensions lying on the principal continuous series, for which $\beta=\sum_{i=1}^{4}\im\lambda_i$.}, but even then it is non-analytic in $\lambda$ and has support only when $\lambda=0$. As observed in~\cite{Arkani-Hamed:2020gyp}, a softening of the UV-behaviour of the amplitude leads to a much better analytic structure for the celestial amplitude, although explicit examples of such amplitudes beyond string theory~\cite{Stieberger:2018edy} are hard to come by.

Note that adding finitely many loop corrections to the Born approximation does not resolve the problem. Focusing on $\omega$-dependence, the sum of all contributions from tree-level to $L$-loops can be written as a finite sum of powers of $\omega$ and $\log\omega$; the Mellin transform of this finite sum is simply the sum of the Mellin transforms of each term. Such a term leads to a Mellin integral of the form
\be\label{Lloop}
\int_{0}^{\infty}\d\omega\,\omega^{\beta+2J-3+m}\,\left(\log\omega\right)^n\,,
\ee
for some $m,n\in\N$. These integrals are generically un-defined, and if $\beta+2J-3+m=\im\,\lambda$ for $\lambda\in\R$ then they are distributional, proportional to a $n$-times differentiated delta function, $\delta^{(n)}(\lambda)$.

However, the situation is different if one considers the resummation of \emph{infinitely} many powers of $\omega$ or $\log\omega$. For instance, the Mellin transforms of 
$$
\e^{-\omega}=\sum_{n=0}^{\infty}\frac{(-\omega)^n}{n!} \qquad \mbox{or} \qquad \frac{1}{1+\omega}=\sum_{n=0}^{\infty}\omega^n\,,
$$
are perfectly analytic in the complex plane, apart from isolated poles -- in other words, they are meromorphic functions of the conformal dimension. For gravitational scattering where $\rg^2=8\pi G$ and 
\be\label{GRBAmp}
\mathrm{M}_{\mathrm{Born}}(s,t)=\frac{8\pi\,G\,s^2}{t}=-\frac{8\pi\,G\,\omega^{2}}{z}\,,
\ee
the celestial amplitude
\be\label{GRCAmp}
\cA_{\mathrm{Born}}(\beta,z)=-\frac{8\pi\,G}{z}\int_{0}^{\infty}\d\omega\,\omega^{\beta+1}\,,
\ee
is non-analytic (indeed, at best distributional in $\beta$ and generally un-defined) due to the UV-incompleteness of the tree-level scattering process. We will now show that a particular resummation, arising naturally from the eikonal approximation, leads to phases of the form
\be\label{epresum}
\sum_{n=0}^{\infty}\frac{(\im\omega)^n\,(\log\omega)^n}{n!}=\omega^{\im\omega}\,,
\ee
whose oscillations restore analyticity in $\beta$ for the resulting celestial eikonal amplitude, apart from an infinite series of isolated poles. First though, we recall some basic facts about eikonal scattering in QFT.


\subsection{Eikonal amplitudes in momentum space}

For $2\to2$ scattering in a relativistic QFT, the \emph{eikonal approximation} refers to considering the small-angle -- or equivalently, large impact parameter -- limit of the scattering amplitude. Suppose the scattered states have incoming momenta $p_1$, $p_2$ and outgoing momenta $p_1'\equiv p_3$, $p_2'\equiv p_4$, respectively; the leading eikonal approximation then corresponds to the kinematic regime where $s\gg-t$ (small angle) and\footnote{From now on, we will set $\hbar=1$, reinstating it only when needed.} $G\,s\gg\hbar$ (trans-Planckian energy), in terms of the Mandelstam variables \eqref{mandelstam}. When the $t/s\to0$ limit isolates ladder and crossed ladder diagrams involving exchanges of the highest-spin particle in the theory (a feature often referred to as `graviton dominance'~\cite{tHooft:1987vrq}) between the scalars at each loop-order -- and these diagrams can be re-summed as an exponential series -- the resulting \emph{eikonal amplitude} is controlled entirely by the single exchange Born term (despite being all-orders in the coupling)~\cite{Torgerson:1966zz,Cheng:1969eh,Abarbanel:1969ek,Cheng:1969tje,Levy:1969cr,Wallace:1973iu,Wallace:1977ae}. 

This phenomenon is referred to as \emph{eikonal exponentiation}. While it has been established in many interesting QFTs (see, for instance, the comprehensive recent review~\cite{DiVecchia:2023frv}) -- including scalars exchanging spin-0 mesons, gravitationally-coupled scalars, QED and scalar QED, and string theory -- it fails for pure $\Phi^3$ theory (due to the presence of diagrams which dominate ladders in the $t/s\to0$ limit~\cite{Tiktopoulos:1971hi,Eichten:1971kd,Kabat:1992pz}) and non-abelian gauge theory (as ladders cannot be trivially re-summed due to the colour structures). There are also interesting cases where eikonal exponentiation is widely-believed to hold, but there is no direct proof at the level of Feynman diagrams, such as infinite-spin point particles coupled to gravity or electromagnetism~\cite{Guevara:2018wpp,Chung:2018kqs,Guevara:2019fsj,Arkani-Hamed:2019ymq,Moynihan:2019bor,Haddad:2021znf,Bianchi:2023lrg,Gatica:2023iws,Luna:2023uwd}. 

Restricting our attention to massless external particles, when eikonal exponentiation holds (in four spacetime dimensions), the resulting eikonal amplitude takes the form\footnote{In this paper, we discard the subtraction of forward scattering from the amplitude; including this would simply result in an additional term proportional to $\delta^{2}(q_\perp)$ in \eqref{geneikamp}.}
\be\label{geneikamp}
\im\,\cM_{\mathrm{eik}}=4\,(p_1\cdot p_2)\,\int \d^{2}x^{\perp}\,\exp\!\left(\im\,\delta_0(x^\perp)-\im\,q_\perp\,x^\perp\right)\,,
\ee
where the integral is over the 2-dimensional space $x^{\perp}$ orthogonal to $p_1$ and $p_2$, and $\delta_0$ is the (leading) eikonal phase:
\be\label{geneikphase}
\delta_0(x^\perp):=\int\frac{\d^4 q}{(2\pi)^2}\,\delta(2p_1\cdot q)\,\delta(2p_2\cdot q)\,\e^{\im\,q\cdot x}\, \mathrm{M}_{\mathrm{Born}}(s,q)\,,
\ee
for $\mathrm{M}_{\mathrm{Born}}$ the Born amplitude for a single $t$-channel exchange (of highest available spin) between the scattered particles.

In many cases, the integrals appearing in \eqref{geneikamp} can be evaluated analytically. For instance, in the case of gravitational scattering of massless scalars, the resulting eikonal amplitude was computed long ago~\cite{tHooft:1987vrq}:
\be\label{tHooftamp}
\cM_{\mathrm{eik}}=-\frac{2\pi\im\,s}{\mu^2}\,\frac{\Gamma(1-\im\,G\,s)}{\Gamma(\im\,G\,s)}\left(\frac{4\,\mu^2}{-t}\right)^{1-\im\,G\,s}\,,
\ee
where $G$ is Newton's constant and $\mu$ is a mass scale acting as an IR regulator. Exploiting the Gamma function identity $\Gamma(1+x)=x\,\Gamma(x)$, this is equivalent to
\be\label{tHooftamp2}
\cM_{\mathrm{eik}}=\underbrace{\frac{8\pi\,G\,s^2}{t}}_{\mathrm{Born}}\,\underbrace{\frac{\Gamma(-\im\,G\,s)}{\Gamma(\im\,G\,s)}\left(\frac{4\,\mu^2}{-t}\right)^{-\im\,G\,s}}_{\mathrm{phase}}\,.
\ee
This illustrates a general feature of eikonal amplitudes: while $\cM_{\mathrm{eik}}$ is all-orders in the coupling, it is actually equal to the classical Born approximation times a phase. When integrating over the full energy scale of this amplitude in a Mellin transform, this phase plays a crucial role.


\subsection{Celestial eikonal amplitudes}

Having established that the Mellin transform of the Born amplitude \eqref{GRBAmp} is ill-defined, one must hope that considering the full eikonal amplitude, including the contribution from the phase in \eqref{tHooftamp2}, renders the Mellin transform finite. In fact, this is precisely the case; this is related to the role of the eikonal phase in partially restoring unitarity of the Born graviton exchange in the high-energy regime of gravitational scattering~\cite{Muzinich:1987in,Amati:1987wq,Amati:1988tn,Amati:1990xe,Giddings:2007bw,Giddings:2009gj,DiVecchia:2023frv}\footnote{See also a recent discussion of the phenomenon in AdS~\cite{Chen:2024iuv}.}.

The small-angle constraint $s\gg-t$ in momentum space corresponds to $z\ll1$ in the parametrization of $\cA(\beta,z)$. However, the celestial analogue of the trans-Planckian energy/classical regime is not immediately clear. In~\cite{deGioia:2022fcn}, it was argued that the celestial version of this high-energy regime corresponds to large conformal dimensions, $\beta\gg1$. We will confirm that this is indeed the case below, but for now we simply define the celestial eikonal amplitude as the Mellin transform of \eqref{tHooftamp2}, with the understanding that the small $\omega$ region of the integral will generically receive quantum corrections. In any case, the problematic portion of the perturbative Mellin transform was the UV, or large $\omega$, region of integration. 

Thus, the dynamical part of the 4-point celestial eikonal amplitude for $2\to2$ gravitational scattering of massless scalars is obtained by taking the appropriate part of \eqref{tHooftamp2} with the kinematics indicated by \eqref{AAmp}:
\begin{equation}
     \mathcal{A}_{\text{eik}}(\beta, z) = -\frac{G}{z}\int_0^{\infty} \d\omega \, \omega^{\beta+1}\,  \frac{\Gamma(-\mathrm{i} G\,\omega^2)}{\Gamma(\mathrm{i} G\,\omega^2)} \left( \frac{4\, \mu^2}{\omega^2\, z}\right)^{-\mathrm{i}\, G\, \omega^2} \, . \label{eq:Aeik4}
\end{equation}
An interesting feature of this celestial eikonal amplitude can be observed even before establishing the convergence properties of the Mellin transform. By rescaling $\omega\to G^{-1/2}\omega$ all dependence on $G$ can be factorized out from the amplitude:
\be\label{Aeikfact}
\cA_{\mathrm{eik}}(\beta,z)=\frac{-1}{2\,G^{\beta/2}\,z}\int_{0}^{\infty}\d\omega\,\omega^{\beta/2}\frac{\Gamma(-\mathrm{i}\,\omega)}{\Gamma(\mathrm{i}\,\omega)} \left( \frac{4\, \tilde\mu^2}{\omega\, z}\right)^{-\mathrm{i}\, \omega}\,,
\ee
after redefining $\mu$ to the dimensionless quantity
\begin{equation}
    \tilde \mu^2 := G\, \mu^2\,.
\end{equation}
In momentum space, $G$ served as a scale controlling the order of exchange multiplicity in the ladder diagram expansion of the eikonal approximation. Upon integrating out the energy scale dependence in the Mellin transform, such coupling scales will inevitably be lost. The celestial eikonal amplitude is thus manifestly non-perturbative in $G$ (going like $G^{-\beta/2}$), and the small $G$-expansion from momentum space is replaced by the analytic structure of $\cA_{\mathrm{eik}}$ in $\beta$ -- provided the Mellin transform is in fact well-defined.

Now, the Born approximation of this celestial amplitude was exponentially divergent in large $\omega$, so to establish the convergence of the Mellin transform, we expand the integrand of \eqref{Aeikfact} at large $s$. Using the Stirling approximation
\be\label{Stirling}
    \Gamma(x)= \sqrt{\frac{2\pi}{x}}\left(\frac{x}{\e}\right)^x \left(1+O(x^{-1})\right) \, ,
\end{equation}
valid for $|\mathrm{arg}\,x|<\pi$, we get the useful formula
\begin{equation}
    \frac{\Gamma(-\i\, x)}{\Gamma(\i\, x)}=\im\, \e^{2\i\,x }\,x^{-2\i\, x}\left(1+O(x^{-1})\right)\,,
\end{equation}
and it follows that the integral that controls the convergence of the Mellin transform \eqref{Aeikfact} is
\begin{equation}\label{eikUV}
\int_{0}^{\infty}\d\omega\,\omega^{\beta/2}\,\omega^{-\im\,\omega}\,
\exp\!\left[
\im\omega\,
\left(2+\log\left(\tfrac{z}{4\,\tilde\mu^2}\right)
\right)
\right]
\, .
\end{equation}
Thus, the question of whether or not the celestial eikonal amplitude $\cA_{\mathrm{eik}}(\beta,z)$ is well-defined boils down to considering Mellin integrals of the form
\be\label{Mellinmaster}
\int_{0}^{\infty}\d\omega\,\omega^{a-1}\,\e^{\im\,b\,\omega}\,\omega^{-\im\,c\,\omega}\,,
\ee
for $a,b\in\C$ and $c\in\R_+$ some constants which do not depend on $\omega$. We will now establish that these integrals are well-defined and analytic in $a$, and explore the resulting analyticity properties of the celestial eikonal amplitude. Further details are provided in appendix~\ref{appendix:eik_gamma}, where we study the properties of an analytically continued version of~\eqref{Mellinmaster} which we call the eikonal Gamma function.


\subsection{Analyticity of the celestial eikonal amplitude}\label{subsec:4pt_analyticity}

\begin{figure}
    \centering
    \includegraphics{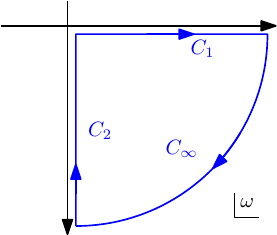}
    \caption{Contour of integration for the analytically-continued Mellin integral.}
    \label{fig:contour-eiko-gamma}
\end{figure}

To establish that the Mellin integral \eqref{Mellinmaster} is well-defined and analytic in $a$, one follows exactly the same line of argument that is used to analytically continue the Gamma function. Take the integrand of \eqref{Mellinmaster}, and integrate it on the contour shown in Figure~\ref{fig:contour-eiko-gamma}.
On the lower right quadrant of the $\omega$-plane, $\exp(-\im c \omega\,\log\omega)$ is exponentially suppressed for $c>0$ and the contribution of $C_\infty$ vanishes. By Cauchy's theorem, this leaves us with $\int_{C_1}+\int_{C_2}=0$. The integral on $C_2$ is given by
\begin{equation}
   -(-\im)^a\,\int_0^\infty \d\omega\, \omega^{a-1}\,\exp\!\left[\left(b+\frac{\pi}{2}\,c\right)\omega\right]\,\omega^{-c\,\omega}\,,
\end{equation}
which is obviously well-defined for $c>0$, and analytic in $a\in\C$ apart from poles at negative integers (which we discuss below). This guarantees that the integral on $C_1$ -- the Mellin integral \eqref{Mellinmaster} itself -- is perfectly well-defined and analytic (apart from isolated poles in $a$). Therefore, the gravitational celestial eikonal amplitude $\cA_{\mathrm{eik}}(\beta,z)$ is well-defined and analytic, apart from potential poles in $\beta$, which we now consider.

\paragraph{Remark:} It is worth making the following observation, which we will elaborate upon later, in Section~\ref{sec:subleading_eikonal}. The oscillatory integral \eqref{Mellinmaster} is defined on a contour $C_1$ which hoovers $\epsilon$-away from the real axis, and thus is actually defined with an $\im\epsilon$ regulator:
\be
\int_{0}^{\infty}\d\omega\,\omega^{a-1}\,\e^{\im\,b\,\omega}\,\omega^{-\im\,(c\,\omega-\im\,\epsilon)}\,,
\ee
The dependence on $\epsilon$ drops out of the answer, rendering its inclusion a formality. However, we will see later that the inclusion of  inelastic effects linked to the emission of gravitational radiation, which are unavoidably present in gravity, \emph{automatically} leads to an $\im\epsilon$ regulator for celestial eikonal amplitudes. This suggests a possible link between analyticity in $\beta$, causality and soft radiation for celestial amplitudes.


\subsubsection{Poles in $\beta$}
Having established that the portion of the integral at large $\omega$ is convergent, we can now study the analytic structure in the $\beta$ plane. As discussed before, one should keep in mind that analytic properties related to the small $\omega$ region of the Mellin integral \emph{will} receive quantum corrections, as this region of integration is outside of the classical regime of the eikonal approximation.   

In order for our discussion to remain generic, we introduce a generalisation of \eqref{Mellinmaster} above, which also describes the full celestial eikonal amplitude~\eqref{Aeikfact}
\begin{equation}\label{eikGammadef2}
    I(a)
    :=\int_0^\infty \d\omega\, \omega^{a-1}\, f(\omega)\, \omega^{-c\, \omega}\,,
\end{equation}
where $f$ is analytic near $0$ and does not spoil convergence at infinity, $a\in\C$ and $c>0$ is a fixed numerical constant of order $1$.

Firstly, note that it is clear that the large-$\omega$ region of integration in \eqref{eikGammadef2} cannot lead to any divergence. That is, for any $L>0$ fixed, the integral
\begin{equation}
    \int_L^\infty \d\omega\, \omega^{a-1}\, 
    f(\omega)
    \, \omega^{-c\, \omega}\,,
\end{equation}
is analytic everywhere as a function of $a$, since no value of $a$ can overpower the exponential suppression which would render the integral divergent. Thus, any poles in $I(a)$ must come from the integral near the origin:
\begin{equation}
   I_L(a)= \int_0^L \d\omega\, \omega^{a-1}\, f(\omega)\,\omega^{-c\,\omega}\,,
\end{equation}
for $L>0$ arbitrarily small. On the interval $(0,L)$, we can Taylor expand $\omega^{-c \omega}$ and the function $f(\omega):=\sum_{m=0}^\infty f_m \omega^m$ (assumed to be analytic) to give
\begin{equation}\label{EGampole1}
 I_L(a)=   \sum_{m,n=0}^\infty \frac{(-c)^n}{n!}\, f_m\, \int_0^L \d\omega\, \omega^{a+n+m-1}\, (\log\omega)^n\,, 
\end{equation}
and then integrate term-by-term.

Each individual term in \eqref{EGampole1} can be easily integrated, and generates singularities according to
\begin{equation}
    \int_0^{L}\d\omega\, \omega^{a+n+m-1}\, (\log\omega)^n \underset{a\to n}{\sim} \frac{(-1)^n\, n!}{(a+n+m)^{n+1}}+\:{\rm regular}\,.
\end{equation}
Therefore, keeping in mind that all the poles of $I(a)$ come from those of $I_L(a)$, which occur at negative integers, the initial integral \eqref{eikGammadef2} has poles of the form:
\begin{equation}\label{EGampole2}
\begin{split}
 I(a)&\underset{a\to k}{\sim} \sum_{n+m=k} \frac{(-c)^n\, f_m}{(a+k)^{n+1}}+\:{\rm regular} \\
  &\sim \frac{(-c)^k\, f_0}{(a+k)^{k+1}}+\frac{(-c)^{k-1}\, f_1} {(a+k)^{k}}+\dots+\frac{f_k}{(a+k)}+\:{\rm regular}\,.
\end{split}
\end{equation}
Thus, the only singularities are poles located at the negative integers, of increasing order as one moves further along the negative real axis.

For the actual celestial eikonal amplitude~\eqref{Aeikfact}, the $f_n$ coefficients can be extracted from the Taylor expansion of the ratio of Gamma functions, which can be obtained from\footnote{This is obtained from a well known expansion for the Gamma function, see e.g., 5.7.3 of \url{https://dlmf.nist.gov/5.7}.}
\begin{equation}\label{eq:Gamma-Ratio-Zeta}
    \frac{\Gamma(-\mathrm{i}\,\omega)}{\Gamma(\mathrm{i}\,\omega)} = -\exp\!\left(2\mathrm{i}\,\gamma_E\,\omega\right)\, \exp\!\left[ \sum_{k\geq1} \frac{2\,\zeta(2k+1)}{2k+1}\, (\mathrm{i}\,\omega)^{2k+1}\right] \, ,
\end{equation}
where $\gamma_E$ is Euler's constant and $\zeta(k)$ is the Riemann zeta function. 
The full eikonal amplitude can the be re-written as
\begin{equation}
\label{ceik0}
    \cA_{\mathrm{eik}}(\beta,z)= \frac{-1}{2\,G^{\beta/2}\,z}\,\int_{0}^{\infty}\d\omega\,\omega^{\beta/2}\, \e^{{\i b}\,\omega} 
   \, \exp\!\left[ \sum_{k\geq1} \frac{ 2\, \zeta(2k+1)}{2k+1}\,  (\mathrm{i}\, \omega)^{2k+1}\right]
   \omega^{\im\,\omega}\,,
\end{equation}
where
\begin{equation}
\label{eq:CE-def}
    b:=\log\left(\frac{C_E\, z}{4\,\tilde \mu^2} \right)\,,\qquad C_E\equiv \exp(2\gamma_E)\,.
\end{equation}
Note that the appearance of the Euler constant in this expression can be reabsorbed in the definition of the IR regulator by taking $\tilde{\mu}^2\to C_E\,\tilde\mu^2$. From \eqref{ceik0} we can immediately identify
\be\label{fncoeffs}
f_0=-1\,, \quad f_1=\im\,b\,, \quad f_2=-\frac{b^2}{2}\,, \quad f_3=-\im\,b^3-\frac{2\,\im}{3}\,\zeta(3)\,, \quad \ldots\,,
\ee
as the coefficients of the poles in the expansion \eqref{EGampole2}.

\medskip

Putting together all of these pieces, we have established that
\begin{equation}
    \mathcal{A}_{\rm eik}(\beta,z)\underset{\beta\to -2(n+1)}{\sim} -\frac{G^{n+1}}{2z}\,\left(\frac{(-\im)^n}{(\frac{\beta}{2}+n+1)^{n+1}} + \textrm{lower-order poles}\right)+{\rm regular}\,.
\end{equation}
In particular, $\cA_{\mathrm{eik}}(\beta,z)$ has poles at integers
\be
\beta=-2\,(n+1)\,, \qquad n=0,1,2,\ldots\,,
\ee
lying on the negative real axis of the $\beta$-plane. These poles are of increasing order as one moves along the negative real axis:
\be\label{betapoleorders}
\begin{split}
\beta=-2 & \qquad \mbox{simple pole }\,, \\
\beta=-4 & \qquad \mbox{simple and double poles}\,, \\
\beta=-6 & \qquad \mbox{simple, double and triple poles}\,, \\
\vdots & \qquad \vdots \\
\beta=-2n & \qquad \:{\rm all\ poles\ up\ to\ } n^{\mathrm{th}}-\mbox{order}\,.
\end{split}
\ee
As the residues at $\beta=-2n$ are proportional to $G^n$, this indicates that these poles can be viewed as some sort of soft $n$-loop effect, as their origin lies in a $(\omega \log\omega)^n$ contribution to the eikonal amplitude. These residues also include up to $\log(z)^n$ dependence in $z$. Furthermore, the regularity of $\cA_{\mathrm{eik}}(\beta,z)$ on the positive real $\beta$-axis is in accordance with general expectations for a UV-complete celestial amplitudes~\cite{Arkani-Hamed:2020gyp}.

Note also that the residues of the lower-order poles in \eqref{EGampole2} depend on $z$ : they are a mixture of the Taylor expansion of the $\e^{\i b\omega}$ and the exponential containing the Riemann zeta functions. Hence the pole expansion appears to be of uniform transcendentality if we give $b\sim \log(z)$ a transcendental weight $1$: on a pole $\beta=-2n$, the residue of $\frac{1}{(\beta+2n)^m}$ has  weight $2n-m$\footnote{See e.g. \cite{Huang:2016tag,DHoker:2019blr} for recent accounts on transcendentality in string amplitudes, and references therein.}.

Finally, observe that only \emph{odd} zeta values -- which are single-valued~\cite{Brown:2004ugm,Brown:2013gia} -- appear in these expressions, a property which is also shared by closed string amplitudes~\cite{Stieberger:2013wea,Stieberger:2014hba,Schlotterer:2018zce,Vanhove:2018elu,Brown:2019wna} in flat space and curvature corrections to string scattering in AdS~\cite{Alday:2022xwz,Alday:2023mvu,Alday:2024ksp}. This could be related to the closed string worldsheet model of Verlinde and Verlinde~\cite{Verlinde:1991iu} used to reproduce the eikonal scattering at Planckian energies. The closed-string-like aspect of eikonal scattering ultimately stems from the fact that in four spacetime dimensions, the plane transverse to scattering is two-dimensional, just like the worldsheet of the closed string.

Of course, we must reiterate that the entire pole structure that we have discussed here will receive quantum corrections, as it arises from the small $\omega$ region of the Mellin transform.


\subsubsection{Large $\beta$ limit}

The behaviour of $\cA_{\mathrm{eik}}(\beta,z)$ in the limit as $\beta\to\infty$ is closely intertwined with the UV behaviour of the underlying momentum space amplitude~\cite{Stieberger:2018edy,Arkani-Hamed:2020gyp}. In this limit, the Mellin integral \eqref{Aeikfact} will be dominated by a saddle point at large $\omega$, which is consistent with the classical region of applicability for the eikonal approximation. This confirms that the strict eikonal limit for the celestial amplitude is $z\ll1$ and $\beta\gg1$, as demonstrated in~\cite{deGioia:2022fcn}. We obtained the large $\omega$ regime of the integrand above in \eqref{eikUV}, so our aim is to evaluate the integral in
\begin{equation}
\lim_{\beta\to\infty}\frac{-2}{G^{\beta/2}\,z}\int_{0}^{\infty}\d\omega\,\exp\!\left[\frac{\beta}{2}\,\log\omega-\im\,\omega\,\log\omega+\im\,\omega\,C\right]\,,
\end{equation}
using the saddle-point approximation, where 
\be\label{cconstdef}
C:=2-\log\!\left(\frac{4\,\tilde{\mu}^2}{z}\right)\,,
\ee
is simply a constant with respect to $\omega$ and $\beta$. 

To compute the saddle point, let us abbreviate
\be\label{fdef}
f(\omega):=\frac{\beta}{2}\,\log\omega-\im\,\omega\,\log\omega+\im\,\omega\,C\,,
\ee
and then look for a critical point solving
\begin{equation}
    \frac{\d f(\omega)}{\d\omega} = \frac{\beta}{2\,\omega} - \mathrm{i}\,\log\omega +\im\, (C-1) =0\,.
\end{equation}
This gives a saddle point at
\begin{equation}
    \omega_* = \frac{-\mathrm{i}\,\beta}{2 \, W\!\left(\frac{-\mathrm{i}\beta\,\e^{1-C}}{2} \right)} \, ,
\end{equation}
where $W(x)$ is the Lambert $W$-function. Now, as we interested in the saddle at large $\beta$, observe that the asymptotic behaviour of the Lambert $W$ function is
\begin{equation}
    W(x)= \log x -\log(\log x) + o(1)  \, ,
\end{equation}
as $x\to\infty$. Thus, for large $\beta$, the saddle point is approximated by
\begin{equation}
    \omega_* \sim \frac{-\mathrm{i}\,\beta}{2\,\log\!\left( \frac{-\mathrm{i}\beta }{2}\right)} \, ,
\end{equation}
which is at high (trans-Planckian) energy for sufficiently large $\beta$. To complete the saddle point evaluation of the integral, we also required the second derivative of $f(\omega)$ evaluated at the saddle:
\begin{multline}
        \left.\frac{\d^2f(\omega)}{\d\omega^2}\right|_{\omega=\omega_*} =\left.\left(\frac{-\beta}{2\,\omega^2} -\frac{\im}{\omega}\right)\right|_{\omega = \omega_*}=-\frac{2}{\beta}\,\log\!\left(-\frac{\im\beta}{2}\right)\left[\log\!\left(-\frac{\im\beta}{2}\right)+1\right] \\
        \sim  -\frac{2}{\beta}\left[\log\!\left(-\frac{\im\beta}{2}\right)\right]^2 \, ,
\end{multline}
with the last line representing the leading contribution in the large $\beta$ limit.

Putting the pieces together, the saddle point approximation for the celestial eikonal amplitude in the large $\beta$ is
\begin{equation}\label{eiklbsp}
\begin{split}
\mathcal{A}_{\text{eik}}(\beta, z) &\underset{\beta\to\infty}{\sim}-\frac{2}{G^{\beta/2}\,z}\,\sqrt{\frac{2\pi}{|f''(\omega_*)|}}\,\e^{f(\omega_*)} \\
&\underset{\beta\to\infty}{\sim}-\frac{2}{G^{\beta/2}\,z}\,\frac{\sqrt{\pi\,\beta}}{ \log\!\left( \frac{-\mathrm{i}\beta }{2}\right) } \left( \frac{-\mathrm{i}\,\beta }{2\log\!\left( \frac{-\mathrm{i}\beta }{2}\right)}\right)^{\frac{\beta}{2}} \exp\!\left( -\frac{\beta}{2}\right) \, .
\end{split}
\end{equation}
Overall, we get that:
\begin{equation}
    \cA_{\rm eik}\sim \left(\frac{\beta}{\log(\beta)}\right)^{\beta/2}\e^{-\beta/2}\,.
\end{equation}
See more details on this calculation and numerical approximations in appendix~\ref{appendix:eik_gamma}.

This behaviour is significantly softer than that for the modulus square of the amplitude regulated by black-hole inelastic effects found in~\cite{Arkani-Hamed:2020gyp}:
\begin{equation}
    |\cA_{\text{BH}}(\beta,z)|^2\sim \frac{\Gamma(\beta)}{G^\beta}\sim\frac{ \beta^\beta\, \e^{-\beta}}{G^\beta} \gg |\cA_{\rm eik}(\beta,z)|^2\sim \frac{1}{G^\beta}\left(\frac{\beta}{\log(\beta)}\right)^{\beta}\, \e^{-\beta}\,.
\end{equation}
This seems to suggests the rule of thumb that the stronger the suppression of the amplitude, the larger the growth at positive $\beta$. There exists various bounds on how much amplitudes can be suppressed at high energies, and relation to the Cerulus-Martin bound~\cite{Cerulus:1964cjb} for gapped QFTs, see for instance~\cite{Tourkine:2023xtu,Buoninfante:2023dyd,Haring:2023zwu} and~\cite{Chowdhury:2019kaq,Chandorkar:2021viw,Haring:2022cyf} for Regge growth of gravitational scattering in flat space. It would be very interesting to see if these constraints based on locality have a direct formulation in the space of conformal dimensions.


\subsubsection{Asymptotic series representation with differential operators}\label{subsubsec:4pt_pole}

At this point, we have established that the celestial eikonal amplitude is well-defined, and characterised its analytic and asymptotic structure in $\beta$. Although explicitly computing the integral \eqref{Aeikfact} appears still quite difficult, we show that it is possible to obtain a fairly explicit, albeit formal, expression for the celestial eikonal amplitude in terms of an asymptotic series of differential operators acting with respect to $\beta$. 

To do this, we start from the representation~\eqref{ceik0}. Using elementary shifting and differentiating properties of Mellin transforms\footnote{If we use $F(\beta)$ to denote Mellin transform of $f(\omega)$, then additional powers of $\omega$ or $\log(\omega)$ in the integrand shifts/differentiates $F(\beta)$ in the following way: $F(\beta+n) = \int_0^\infty \omega^{\beta-1} f(\omega) \omega^n$ and $\partial_{\beta}F(\beta) = \int_0^\infty \omega^{\beta-1} f(\omega) \log(\omega)^n$.}, we can formally pull out of the integral it's whole non-trivial part, and reduce it to an differential operator acting on 
\begin{equation}
     \mathcal{F}(\beta, z) := \frac{\mathrm{i}^{\frac{\beta+2}{2}}}{\left[\log\!\left( \frac{C_E \, z}{4\, \tilde{\mu^2}}\right)\right]^{\frac{\beta+2}{2}}}\, \Gamma\!\left( \frac{\beta+2}{2}\right) \, , \label{eq:calF4}
\end{equation}
where $C_E$ was defined in~\eqref{eq:CE-def}, as
\begin{equation}
        \mathcal{A}_{\text{eik}}(\beta, z) = \frac{-1}{2\,G^{\beta/2}\,z}\, \exp\!\left[2 \mathrm{i}\,  \e^{2\partial_\beta} \, \partial_{\beta}\right] \exp\!\left[ \sum_{k\geq1} \frac{2\,\zeta(2k+1)}{2k+1} \left( \mathrm{i}\, \e^{2\partial_\beta}\right)^{2k+1}\right] \, \mathcal{F}(\beta, z) \, , \label{eq:Abeta_eik}
\end{equation}
The operator $\e^{a\,\partial_{\beta}}$ acts on functions of $\beta$ by shifting 
\be\label{shiftbeta}
\beta\to\beta+a\,.
\ee
Note that the derivative operators $\partial_{\beta}$ and $\e^{\partial_{\beta}}$ commute. 

The expression \eqref{eq:Abeta_eik} is remarkably compact, and manifests the celestial eikonal amplitude as a collection of shift operators in $\beta$ acting on the conformal primary \eqref{eq:calF4}. However, this expression must be viewed as an \emph{asymptotic series}, approximating the true celestial eikonal amplitude in terms of a series expansion of the various exponentials. We demonstrate the asymptotic nature of this series explicitly in Appendix~\ref{appendix:eik_gamma}, but emphasize that even with its formal nature, \eqref{eq:Abeta_eik} captures many interesting properties of the celestial eikonal amplitude. For instance, \eqref{eq:Abeta_eik} captures the divergent nature of the Born approximation: setting $\beta=-2$ to isolate the $O(G)$ contribution produces a singularity from the Gamma function in $\mathcal{F}(\beta,z)$.

Perhaps more remarkably, the asymptotic series also encodes the pole structure of $\cA_{\mathrm{eik}}(\beta,z)$ in the $\beta$-plane (up to quantum corrections). To see this, observe that the Gamma function appearing in \eqref{eq:calF4} has simple poles appearing at 
\be\label{betapoles}
\beta=-2\,(n+1)\,, \qquad n\in\N\,,
\ee
lying on the negative real axis of the $\beta$-plane. The residues of $\mathcal{F}(\beta,z)$ at these simple poles are
\be\label{betaresidues}
\left.\mathrm{Res}\!\left(\mathcal{F}(\beta,z)\right)\right|_{\beta=-2(n+1)}=\frac{(-1)^{n}\,\im^{-n}}{n!}\,\left[\log\!\left(\frac{C_E\,z}{4\,\tilde{\mu}^2}\right)\right]^{n}\,,
\ee
but in \eqref{eq:Abeta_eik} we must also account for the action of the exponentiated differential operators. Each of the two exponential differential operators acting on $\mathcal{F}$ in \eqref{eq:Abeta_eik} contains $\e^{2\partial_{\beta}}$; as noted before, these have the effect of shifting the argument of $\cF$ -- and of the Gamma function in particular -- by two. Therefore, expanding each of the exponential operators in \eqref{eq:Abeta_eik} results in an infinite (asymptotic) series of terms, each of which has poles on the negative real axis which are shifted ever further to the right. 

In addition, each of these `shifted' terms in the expansion is also accompanied by the action of a (non-exponentiated) differential operator $\partial_{\beta}^{k}$, for $k=1,2,\ldots$. For simplicity, consider the terms containing only a single such derivative (i.e., $k=1$); these will act on Gamma functions whose arguments have been shifted \emph{at least} once, so the minimal case is
\be\label{hpgamma1}
\partial_{\beta}\Gamma\!\left(\frac{\beta}{2}+2\right)=\frac{1}{2}\,\Gamma\!\left(\frac{\beta}{2}+2\right)\,\psi^{(0)}\!\left(\frac{\beta}{2}+2\right)\,,
\ee
where $\psi^{(0)}$ is the digamma function. Crucially, $\psi^{(0)}$ has simple poles wherever its argument is equal to a negative integer, so the combination $\partial_{\beta}\Gamma(\beta/2+2)$ has a \emph{double pole} at $\beta=-4$, as desired. 

Generalizing this argument to account for all possible poles arising in the asymptotic series indicates that $\cA_{\mathrm{eik}}(\beta,z)$ will have a maximal pole of order $n$ at $\beta=-2n$ on the negative real axis, as well as lower order poles. It is somewhat surprising to see that \eqref{eq:Abeta_eik} -- which is only an asymptotic series for the true celestial eikonal amplitude -- nevertheless correctly predicts its pole structure.


\section{Celestial eikonal-probe amplitudes in curved space}\label{sec:2pt_celestial}

Having established the existence of celestial eikonal amplitudes for $2\to2$ gravitational scattering of massless scalars, we now turn to a closely related problem: the probe scattering of a massless scalar in a stationary, curved spacetime at large distances from sources. These $1\to1$ \emph{eikonal-probe} amplitudes are closely related -- but not equal to -- standard $2\to2$ eikonal amplitudes in momentum space, and have many basic `eikonal' properties. They are controlled by an eikonal phase, are equal to a Born amplitude (for tree-level scattering between the probe and spacetime source) times a phase and contain all orders in $G$.

In this section, we show that celestial eikonal-probe amplitudes are also well-defined, manifestly non-perturbative objects and explore their analytic and asymptotic properties. This gives a perturbation exact celestial two point function on curved spacetimes. We also establish a dictionary for mapping between celestial eikonal amplitudes and the celestial eikonal-probe amplitude on a gravitational shockwave, and compare with the analogous story in electromagnetism.


\subsection{Eikonal-probe scattering in momentum space}

In the eikonal approximation of $2\to 2$ scattering, it is clear that each particle sees the other as a fixed, classical object. This suggests that it should be possible to provide an alternative description of eikonal scattering in terms of background field scattering: namely, the $1\to 1$ probe scattering of one of the particles in a fixed classical background, which is treated non-perturbatively and sourced by the other particle, evaluated at large impact parameter. We refer to such a background field amplitude as an \emph{eikonal-probe} amplitude; more precisely, in the context of gravitational scattering:
\begin{defn}[Gravitational eikonal-probe amplitude]
Consider any asymptotically flat, stationary solution of the Einstein equations with a spatially compact source. An \emph{eikonal-probe} amplitude on this background is the $1\to1$ tree-level scattering amplitude of a gravitationally-coupled probe particle, evaluated at large impact parameter with respect to the source of the spacetime. 
\end{defn}
The basic intuition is that there should be a relationship between eikonal scattering of two particles and the eikonal-probe scattering of one particle in the background of the other.

This was first made precise for the example \eqref{tHooftamp} of gravitational scattering of massless scalars by 't Hooft~\cite{tHooft:1987vrq}, who showed that $\cM_{\mathrm{eik}}$ is proportional to the eikonal-probe amplitude for a massless scalar on the Aichelberg-Sexl shockwave metric sourced by the other massless scalar~\cite{Aichelburg:1970dh,Dray:1984ha}. This correspondence was subsequently developed for eikonal electromagnetic and gravitational scattering of massive scalars, but in the non-relativisitic language of potential scattering~\cite{Jackiw:1991ck,Kabat:1992tb}. More recently, it was shown that eikonal-probe amplitudes \emph{always} have the structure of an eikonal amplitude~\cite{Adamo:2021rfq}. Furthermore, in this work it was conjectured that when the source of the background spacetime has (at least, at leading order in $G$) a particle-like interpretation, this eikonal-probe amplitude encodes the true QFT eikonal amplitude for $2\to2$ scattering between the probe and the metric source.

The precise correspondence is as follows: given a stationary, asymptotically flat metric $g$ with spatially compact source and ADM mass $M$, consider a probe particle (for our purposes, this will always be a massless scalar) of initial momentum $p$ and final momentum $p'$ on this spacetime. Let $M_2$ denote the eikonal-probe scattering amplitude of this particle. Then~\cite{Adamo:2021rfq}
\be\label{eikprobecon}
M_{2}=\frac{\pi\,\delta(p'_0-p_0)}{2\,M}\,\cM_{\mathrm{eik}}\Big|_{\substack{p_1=p \\
p_2=P}}:=\frac{\pi\,\delta(p'_0-p_0)}{2\,M}\,\cM_{\mathrm{eik}}^{g}\,,
\ee
where $p_0$, $p'_0$ denote the time-like components of the incoming/outgoing probe momentum, $P$ is the time-like four-momentum of the source of the metric and $\cM_{\mathrm{eik}}$ is the $2\to2$ eikonal amplitude for gravitational scattering between the probe and the massive source. It will be convenient to denote $\cM_{\mathrm{eik}}|_{p_1=p,p_2=P}$ by $\cM_{\mathrm{eik}}^{g}$, which is a quantity that is \emph{a priori} defined with respect to eikonal-probe setup (making reference only to the probe momentum and background metric $g$). This relation also extends to ultrarelativisic limits of stationary, asymptotically flat spacetimes, where it is slightly modified to~\cite{Adamo:2021rfq,Adamo:2022rob}
\be\label{eikprobeUR}
M_{2}=\frac{\pi\,\delta(p'_+-p_+)}{4\,P_-}\,\cM_{\mathrm{eik}}\Big|_{\substack{p_1=p \\
p_2=P}}:=\frac{\pi\,\delta(p'_+-p_+)}{4\,P_-}\,\cM_{\mathrm{eik}}^{g}\,,
\ee
where the lightfront coordinates $x^{\mu}=(x^-,x^\perp,x^+)$ are chosen such that the source is localized on a constant $x^-$ lightfront, with lightfront energy $P_-$.

Eikonal-probe scattering amplitudes have been computed explicitly for scalar scattering on Schwarzschild and Kerr spacetimes as well as the the Aichelberg-Sexl shockwave. By virtue of the correspondences \eqref{eikprobecon}, \eqref{eikprobeUR}, these eikonal-probe amplitudes are also controlled by an eikonal phase $\delta_0$, just like the full $2\to2$ eikonal amplitudes. While these formulae have been computed for generic probe masses~\cite{Kabat:1992tb,Adamo:2021rfq}, our interest will be in those cases where the probe scalar is massless.

The eikonal-probe amplitude on an Aichelberg-Sexl gravitational shockwave background is immediately identified by taking $\cM_{\mathrm{eik}}^{\mathrm{shock}}$ to be \eqref{tHooftamp} subject to the identification of the Mandelstam invariants $s=p_+\,P_-$ and $t=-(p-p')_{\perp}^2$; the corresponding eikonal phase is~\cite{tHooft:1987vrq}
\be\label{epshock}
\delta_0^{\mathrm{shock}}(x^\perp)=-4G\,p_+\,P_-\,\log\!\left(\mu\,|x^{\perp}|\right)\,.
\ee
For a massless scalar scattering on the Schwarzschild metric, the eikonal-probe amplitude is encoded by~\cite{Kabat:1992tb,Adamo:2021rfq}
\be\label{eikprobeSchw}
\cM_{\mathrm{eik}}^{\mathrm{Schw}}=-32\pi\,G\,\frac{(p\cdot P)^2}{(p-p')_{\perp}^2}\,\frac{\Gamma(-2\im\,G\,p\cdot P)}{\Gamma(2\im\,G\,p\cdot P)}\left(\frac{4\,\mu^2}{(p-p')_{\perp}^2}\right)^{-2\im\,G\,p\cdot P}\,,
\ee
where $P_{\mu}=M(1,0,0,0)$ and the corresponding eikonal phase is
\be\label{epSchw}
\delta_0^{\mathrm{Schw}}(x^\perp)=-4\,G\,p\cdot P\,\log\!\left(\mu\,|x^{\perp}|\right)\,.
\ee
Finally, the eikonal-probe amplitude for a massless scalar on the Kerr metric is simply a phase rotation of the Schwarzschild result~\cite{Adamo:2021rfq}:
\be\label{eikprobeKerr}
\cM_{\mathrm{eik}}^{\mathrm{Kerr}}=\e^{\im\,(p-p')_{\perp}\,a^{\perp}}\,\cM_{\mathrm{eik}}^{\mathrm{Schw}}\,,
\ee
where $a^{\perp}$ are the components of the (mass-rescaled) covariant spin vector
\be\label{Kerrspin}
a^{\mu}:=\frac{1}{2\,M^2}\,\epsilon^{\mu\nu\alpha\beta}\,P_{\nu}\,S_{\alpha\beta}\,,
\ee
for $S^{\mu\nu}$ the spin tensor. The corresponding eikonal phase is
\be\label{epKerr}
\delta_0^{\mathrm{Kerr}}(x^\perp)=-4\,G\,p\cdot P\,\log\!\left(\mu\,|x^{\perp}-a^{\perp}|\right)\,.
\ee
It should be noted that the massless probe condition dramatically simplifies the resulting eikonal-probe amplitude in this case; for generic mass configurations, the eikonal-probe amplitude on Kerr is a linear combination of confluent hypergeometric functions~\cite{Adamo:2021rfq}. The `eikonal' in eikonal-probe amplitude ensures that these amplitudes are only sensitive to large distance (or linear in $G$, for harmonic coordinates) features of the metric, so the presence of an event horizon does not leave an imprint on \eqref{epSchw} or \eqref{epKerr}.

\medskip

Armed with the known formulae for scalar eikonal-probe amplitudes on the Aichelberg-Sexl gravitational shockwave as well as the Schwarzschild and Kerr black holes, we can now ask what these amplitudes correspond to when Mellin transformed to the celestial sphere. The tree-level celestial Born amplitudes for these processes are typically divergent due to the mixing between UV and IR regimes of the amplitude\footnote{An exception is the Kerr case, where the finite size of the source of the Kerr metric regularizes the Mellin integral of the Born amplitude~\cite{Gonzo:2022tjm,Adamo:2022rob}.}, but we find that the \emph{full} eikonal-probe amplitude is always a finitely defined quantity as the full eikonal amplitude. In particular, like we have seen in section \ref{sec:4pt_celestial} for the $2\to 2$ eikonal case, the exponentiated eikonal phase plays a crucial role in regularizing the Mellin integral.

Beginning with the celestial eikonal-probe amplitude on the gravitational shockwave, we analyse the high-energy behaviour of the amplitude to demonstrate convergence of its Mellin transform, then evaluate the Mellin transform explicitly in terms of certain shift operators acting on the conformal dimensions. In each case, we comment on the analytic structure of the celestial eikonal-probe amplitudes in the space of conformal dimensions.


\subsection{Gravitational shockwave background}\label{subsec:shockwave}

The celestial eikonal-probe amplitude for a massless scalar scattering on any background is obtained by performing a Mellin transform in the frequencies of the incoming and outgoing probe scalars. On the null background of the Aichelberg-Sexl gravitational shockwave, it is easiest to parametrize both the probe momentum $p$, and the momentum of the background source (itself another massless scalar) $P$, in lightfront coordinates. In these variables, we take
\be\label{mmomparam}
p_{\mu}=(p_-,p_\perp,p_+)=\omega\left(|z|^2,\,\frac{z+\bar{z}}{\sqrt{2}},\,-\frac{\im(z-\bar{z})}{\sqrt{2}},\,1\right)\,,
\ee
where $\omega$ is the incoming scalar frequency and $(z,\bar{z})$ labels a point on the celestial sphere. Similarly, for the outgoing scalar probe
\be\label{mmomparam2}
p'_{\mu}=(p'_-,p'_\perp,p'_+)=\omega'\left(|z'|^2,\,\frac{z'+\bar{z}'}{\sqrt{2}},\,-\frac{\im(z'-\bar{z}')}{\sqrt{2}},\,1\right)\,,
\ee
while the momentum of the background source is
\be\label{swbmom}
P_{\mu}=\left(P_-,0,0,0\right)\,,
\ee
where $P_-$ is the lightfront energy of the shockwave. 

The Mandelstam invariants controlling the eikonal-probe amplitude in this parametrization are then given by
\be\label{shockMand}
s=\omega\,P_-\,, \qquad t=-|\omega\,z-\omega'\,z'|^2\,,
\ee
rendering $\cM_{\mathrm{eik}}^{\mathrm{shock}}$ a function of $\omega,\omega'$, the points $(z,\bar{z})$, $(z',\bar{z}')$ on the celestial sphere and the lightfront energy $P_-$ of the background. The overall momentum conserving delta function
\be\label{lfecon}
\delta(p'_+-p_+)=\delta(\omega'-\omega)\,,
\ee
ensures that the incoming and outgoing probe frequencies are equal.

\medskip

We point out at this stage an abuse of notation: in the eikonal-probe context, the variables $z$ and $z'$ are positions on the celestial sphere, \emph{not} conformal cross ratios. The explicit translation to the 4-point kinematics of Section~\ref{sec:4pt_celestial} is:
\begin{equation}
   p_\mu \equiv p_{\mu,1}\,, \quad p'_\mu\equiv p_{\mu,3}\,, \quad z\equiv z_1\,, z'\equiv z_3\,,\quad \omega\equiv \omega_1\,,\quad \omega'\equiv \omega_3\,.
\end{equation}
We have chosen to to work with $z,z'$ in the 2-point context to avoid cluttering notation with additional subscripts. 

Note that here, conformal invariance is broken by the choice of a fixed shockwave background. On the celestial sphere, large $z,z'$ is the limit where incoming $p_1\equiv p$ and outgoing $p_3\equiv p'$ become aligned with the shock, and thus we identify the background as giving rise to an extra, fixed, operator inserted at $z_2\equiv \infty$. This is consistent with general expectations that in celestial holography, eikonal-probe amplitudes will correspond to 2-point functions in a celestial CFT state corresponding to the background, which typically breaks conformal invariance on the sphere\footnote{We thank Eduardo Casali for emphasising this to us.}. This is certainly the case for probe scattering on shockwaves in AdS/CFT~\cite{Cornalba:2006xk,Cornalba:2006xm,Cornalba:2007zb}, and has been confirmed in the celestial context with perturbative calculations~\cite{Pasterski:2020pdk,Gonzo:2022tjm,deGioia:2022fcn}.


\subsubsection{High-energy behaviour and analyticity of the Mellin transform}

The Born amplitude underlying this eikonal-probe amplitude, identified from \eqref{tHooftamp2} as
\be\label{Bornshock}
\cM_{\mathrm{Born}}^{\mathrm{shock}}=\frac{8\pi\,G\,s^2}{t}=-\frac{8\pi\,G\,P_-\,\omega^2}{|\omega z-\omega' z'|^2}\,,
\ee
again leads to generically divergent (or at best non-analytic) Mellin transforms. However, the full eikonal-probe amplitude, including the contribution from the phase in \eqref{tHooftamp2} resolves this divergence in much the same way that it did for the 4-point eikonal amplitude. 

Using the Stirling approximation \eqref{Stirling} for the eikonal amplitude, but now evaluated in the probe kinematics \eqref{shockMand} (with $\omega=\omega'$), gives the high-energy behaviour:
\begin{equation}\label{shockUV}
\lim_{s\to\infty}\cM_{\mathrm{eik}}^{\mathrm{shock}}\sim\mathcal{M}^{\mathrm{UV}}_{\text{eik}}= \frac{P_-^2}{|z-z'|^2} \left( \frac{\e^2\,|z-z'|^2}{4G^2\, P_-^2 \,\mu^2}\right)^{\mathrm{i}G\, \omega P_-}  \, .
\end{equation}
Crucially, we see that the Born prefactor is $\omega$-independent, while the phase component is oscillatory in $\omega$. Indeed, this phase tunes the high-energy behaviour of the amplitude, leading to a manifestly meromorphic (in $\Delta+\Delta'$) Mellin integral at large $s$:
\begin{equation}\label{eq:M2approx}
    \int_0^{\infty} \d\omega \, \omega^{\Delta+\Delta'-2}\,\mathcal{M}^{\text{UV}}_{\text{eik}} =\frac{P_-^2}{|z-z'|^2}\,\frac{\mathrm{i}^{\Delta+\Delta'-1}\,\Gamma(\Delta+\Delta'-1)}{\left(G\,P_-\, \log\left(\frac{\e^2\,|z-z'|^2}{4G^2\, P_-^2 \, \mu^2}\right) \right)^{\Delta+\Delta'-1}} \, . 
\end{equation}
Thus, if the Mellin transform of the high-energy eikonal-probe amplitude is a meromorphic function of the conformal dimensions, we expect that the Mellin transform of the full eikonal-probe amplitude will be as well (in contrast to the Born approximation).


\subsubsection{The celestial eikonal-probe amplitude}

Now, the \emph{celestial} eikonal-probe amplitude is obtained by Mellin transforming with respect to the incoming and outgoing probe frequencies:
\begin{equation}
    \begin{split}
        \widetilde{M}^{\mathrm{shock}}_2 =& \int_0^\infty \d\omega \,\int_0^\infty \d\omega' \, \omega^{\Delta-1}\,(\omega')^{\Delta'-1} \,\frac{\pi\,\delta(\omega'-\omega)}{4\, P_-}\,\mathcal{M}^{\mathrm{shock}}_{\text{eik}} \\
        =\,&\frac{-2\pi^2}{(GP_{-})^{\Delta+\Delta'-2}\,|z-z'|^2} \int_0^{\infty} \d\omega\, \omega^{\Delta+\Delta'-2}\,\frac{\Gamma(-\im\,\omega)}{\Gamma(\im\,\omega)}\left(\frac{4\,\tilde{\mu}^2}{\omega^2\,|z-z'|^2}\right)^{-\im\,\omega} \\
        =\,&\frac{2\pi^2}{(GP_{-})^{\Delta+\Delta'-2}\,|z-z'|^2} \int_0^{\infty} \d\omega\, \omega^{\Delta+\Delta'-2}\, \exp\!\left[2\mathrm{i}\,\gamma_E\, \omega - \mathrm{i}\,\omega\, \log\!\left(\frac{4\,\tilde{\mu}^2}{|z-z'|^2}\right)\right]\\
        &\quad\quad\times \exp\!\left[ \sum_{k\geq1} \frac{ \, 2\,\zeta(2k+1)}{2k+1}\,  (\mathrm{i}\, \omega)^{2k+1}\right] \, \exp\!\left[\mathrm{i}\,\omega\, \log(\omega^2) \right]  \, , \label{eq:tildeM2}
    \end{split}
\end{equation}
where all dependence on the scales $G$ and $P_-$ has been scaled out of the integral and the new IR regulator is defined as $\tilde{\mu}^2:=G^2\,P^2_-\,\mu^2$. This integral is precisely of the form \eqref{Mellinmaster}, whose analyticity properties we have already established. Therefore, the celestial eikonal-probe amplitude \eqref{eq:tildeM2} on a shockwave background is a meromorphic function of $\Delta+\Delta'$.

Using basic properties of the Mellin transform, this integral can be written as an asymptotic series in terms of exponentiated differential operators. Defining:
\begin{equation}
\begin{split}
    \mathcal{F}(\Delta+\Delta') &:= \int_0^{\infty} \d\omega\, \omega^{\Delta+\Delta'-2}\, \exp\!\left[2\mathrm{i} \,\gamma_E \,\omega  - \mathrm{i}\,\omega\, \log\!\left(\frac{4\,\tilde{\mu}^2}{|z-z'|^2}\right)\right] \\
&=\frac{\mathrm{i}^{\Delta+\Delta'-1}\,\Gamma(\Delta+\Delta'-1)}{\left[\log\!\left(\frac{C_E \,|z-z'|^2}{4\,\tilde{\mu}^2}\right)\right]^{\Delta+\Delta'-1}} \, , \label{eq:calF2}
\end{split}
\end{equation}
for $C_E\equiv\e^{2\gamma_E}$, \eqref{eq:tildeM2} becomes
\begin{multline}\label{2pt_function}
    \widetilde{M}^{\mathrm{shock}}_2 =\frac{2\pi^2 }{(GP_-)^{\Delta+\Delta'-2}\,|z-z'|^2}\,\exp\!\left[2\mathrm{i}\,\e^{\partial_{\Delta+\Delta'}} \, \partial_{\Delta+\Delta'} \right]
    \\ \times\,  \exp\!\left[\sum_{k\geq1} \frac{ 2\,\zeta(2k+1)}{2k+1}\,  (\mathrm{i}\, \e^{\partial_{\Delta+\Delta'}} )^{2k+1}\right] \mathcal{F}(\Delta+\Delta') \, ,
\end{multline}
where the operator $\e^{\partial_{\Delta+\Delta'}}$ acts on functions of $\Delta+\Delta'$ by shifting $\Delta+\Delta'\to\Delta+\Delta'+1$.

We emphasize that the representation \eqref{2pt_function} must be viewed as a formal asymptotic series for the celestial eikonal-probe amplitude on a shockwave spacetime. Yet, just as for the 4-point celestial eikonal amplitude, this asymptotic series captures many important features of the celestial eikonal-probe amplitude. As expected, \eqref{2pt_function} is meromorphic in $\Delta+\Delta'$ and encodes the divergent Born approximation, which corresponds to $\Delta+\Delta'=1$.


\subsubsection{Analytic and asymptotic properties}

Having obtained the celestial eikonal-probe amplitude \eqref{eq:tildeM2}, we now consider some of its basic properties, starting with its analytic behaviour in the space of conformal dimensions. Given the similarity in the Mellin integrals with the full eikonal case \eqref{eq:Abeta_eik}, the discussions will largely follow from the pole structure analysis in sec \ref{subsubsec:4pt_pole}; to avoid repetition, we capture the essence of that discussion here. Near $\omega\sim 0$, \eqref{eq:tildeM2} reduces to
\begin{equation}
    \int_0^L \d\omega\, \omega^{\Delta+\Delta'-2}\, \e^{\im\omega\,\log(\omega^2)}\,,
\end{equation}
where $L>0$ is arbitrarily close to $0$. This is easily evaluated by Taylor expanding the exponential 
\begin{equation}
    \sum_{n=0}^\infty \frac{\im^n}{n!}\int_{0}^L\d\omega\,\omega^{\Delta+\Delta'-2+n}\,(\log\omega)^n  \,,
\end{equation}
and each term evaluated as $L\to 0$ to reveal the pole structure in $\Delta+\Delta'$. The $n^{\mathrm{th}}$ term gives:
\begin{equation}
   \lim_{L\to 0} \int_0^{L}\d\omega\, \omega^{\Delta+\Delta'+n-2}\, (\log\omega)^n  = \frac{(-1)^n\, n!}{(\Delta+\Delta'-1+n)^{n+1}}+{\rm regular}\,,
\end{equation}
corresponding to a pole at 
\begin{equation}\label{delres1}
    \Delta+\Delta'-1=-n \quad\text{of order } n+1\,, 
\end{equation}
with residue $\frac{(\im GP_-)^n}{n!}$. Additional, lower-order poles at the same location are generated by Taylor expanding the exponential with odd zeta values in \eqref{eq:tildeM2}. 

Accounting for additional overall factors in the full celestial eikonal-probe amplitude, we have that
\be\label{ceppoles1}
\widetilde{M}_2^{\mathrm{shock}}\underset{\Delta+\Delta'\to-n}{\sim}\frac{2\pi^2\,(GP_{-})^n}{|z-z'|^2}\,\left(\frac{(-\im)^{n-1}}{(\Delta+\Delta'+n)^{n}}+\mbox{ lower-order poles }\right)+\mbox{ regular }\,.
\ee
In particular, $\widetilde{M}_2^{\mathrm{shock}}$ has poles of increasing order at integer values along the negative real axis of the $(\Delta+\Delta')$-plane, and is regular on the $\mathrm{Re}(\Delta+\Delta')>0$ half-plane, features associated with a UV-complete scattering process~\cite{Stieberger:2018edy,Arkani-Hamed:2020gyp}. The $n^{\mathrm{th}}$-order pole at $\Delta+\Delta'=-n$ has a residue proportional to $G^{n}$, indicating that it arises as a sort of soft $n$-loop effect. We also note that the asymptotic series \eqref{2pt_function} also captures the correct pole structure of the celestial eikonal-probe amplitude, in much the same fashion as in the 4-point eikonal amplitude.

\medskip

We can also comment on the asymptotic behaviour of $\widetilde{M}^{\mathrm{shock}}_2$ in the limit of large conformal dimensions: $\Delta+\Delta'\to\infty$; the analysis then follows the same lines as the saddle-point arguments for the large $\beta$ limit of the 4-point celestial eikonal amplitude. In particular, the large $\Delta+\Delta'$ behaviour of the amplitude will be controlled by a saddle-point approximation of the large-$\omega$ regime of \eqref{eq:tildeM2}. 

In other words, we want to approximate the integral
\be\label{epsadd1}
\int_{0}^{\infty}\d\omega\,\exp\!\left[(\Delta+\Delta'-2)\,\log\omega+\im\,\omega\,B\right]\,,
\ee
where 
\be\label{Bdef}
B:=\frac{e^2\,|z-z'|^2}{4\,\tilde{\mu}^2}\,.
\ee
In the limit of interest, there is no difference between $\Delta+\Delta'$ and $\Delta+\Delta'-2$, and the integral is dominated by a saddle at
\be\label{epsadd2}
\omega_*=\im\,\frac{\Delta+\Delta'}{B}\,,
\ee
leading to the approximation
\be\label{epsadd3}
\widetilde{M}^{\mathrm{shock}}_2\underset{\Delta+\Delta'\to\infty}{\sim}\frac{P_-^{2}\,\sqrt{2\pi}}{(GP_-)^{\Delta+\Delta'}\,|z-z'|^2}\,\frac{\sqrt{\Delta+\Delta'}}{\log\!\left(\frac{e^2\,|z-z'|^2}{4\,\tilde{\mu}^2}\right)}\left[\frac{\im\,(\Delta+\Delta')}{\log\!\left(\frac{e^2\,|z-z'|^2}{4\,\tilde{\mu}^2}\right)}\right]^{\Delta}\,\e^{-(\Delta+\Delta')}\,,
\ee
for the large-$(\Delta+\Delta')$ growth of the celestial eikonal-probe amplitude on the shockwave background.


\subsection{Black-hole backgrounds}\label{subsec:black_hole}

The computation of celestial eikonal-probe amplitudes on Schwarzschild and Kerr black holes proceeds along similar lines to the gravitational shockwave case. In both cases, the eikonal phase improves the UV behaviour of the amplitude, ensuring that the Mellin transforms are well-defined\footnote{The Mellin transform of the Born approximation to the Schwarzschild amplitude is divergent in the same way as the shockwave, while in the Kerr case the Mellin transform of the Born amplitude is actually finite due to the finite-size of the source of the Kerr metric~\cite{Gonzo:2022tjm}.} by virtue of being in the same class of integrals as the analytically continued eikonal Gamma function. The analytic and asymptotic properties of the resulting celestial eikonal-probe amplitudes are virtually identical to those of the shockwave, and formal expressions for the amplitudes as asymptotic series of differential operators can also be obtained.

We emphasize that although these celestial eikonal-probe amplitudes are not sensitive to the event horizon of the background black hole, they provide a result which \emph{any} celestial dual for asymptotically flat Schwarzschild or Kerr black holes must reproduce. In particular, one expects probe amplitudes on a black hole to correspond to some two-point function in a dual description on the celestial sphere, and the small-angle limit~\cite{Bautista:2021wfy,Adamo:2023cfp} of this correlator must reproduce the celestial eikonal-probe amplitudes given here.

\medskip

In practical terms, the main difference between the black hole and shockwave calculations is that the background now has a time-like (rather than null) momentum associated with its source. It is thus more convenient to consider the parametrization of the null probe and time-like background momenta in Cartesian, rather than lightfront, coordinates. In these variables, we have
\be\label{bhmpar}
p_{\mu}=(p_0,p_x,p_y,p_z)=\frac{\omega}{\sqrt{2}}\left(1+|z|^2,\,z+\bar{z},\,-\im(z-\bar{z}),\,1-|z|^2\right)\,, 
\ee
for the massless incoming probe,
\be\label{bhmpar2}
p'_{\mu}=\frac{\omega'}{\sqrt{2}}\left(1+|z'|^2,\,z'+\bar{z}',\,-\im(z'-\bar{z}'),\,1-|z'|^2\right)\,,
\ee
for the outgoing probe and
\be\label{bhmpar3}
P_{\mu}=\left(M,\,0,\,0,\,0\right)\,,
\ee
for the black hole source momentum, where $M$ is the mass of the black hole. The associated Mandelstam invariants for the eikonal-probe amplitudes then become
\be\label{BHMand}
s=\omega\,\left(1+|z|^2\right) M+M^2\,, \qquad t=-|\omega\,z-\omega'\,z'|^2\,,
\ee
with the angular dependence in $s$ appearing as a result of the fact that the background momentum is time-like.


\subsubsection{Schwarzschild}

The eikonal-probe amplitude for a massless scalar on the Schwarzschild black hole is given by \eqref{eikprobeSchw}, which can be rewritten as
\be\label{epSchw1}
\cM_{\mathrm{eik}}^{\mathrm{Schw}}=\frac{8\pi\,G\,\alpha(s)^2}{t}\,\frac{\Gamma(-\im\,G\,\alpha(s))}{\Gamma(\im\,G\,\alpha(s))}\,\left(\frac{4\,\mu^2}{-t}\right)^{-\im\,G\,\alpha(s)}\,,
\ee
where this expression is evaluated on the kinematics \eqref{BHMand}, $\mu$ is once again an IR regulator, and
\be\label{alphadef}
\alpha(s):=s-M^2=\omega\left(1+|z|^2\right)\, M\,.
\ee
It is clear that in the high-energy limit $s>>|t|$, this amplitude scales identically to the shockwave example considered above. In particular, this means that the Born amplitude
\be\label{BornSchw}
\cM^{\mathrm{Schw}}_{\mathrm{Born}}=\frac{8\pi\,G\,\alpha(s)^2}{t}\,,
\ee
has a divergent Mellin transform (as shown in~\cite{Gonzo:2022tjm}), but the inclusion of the phase in \eqref{epSchw1} serves to regulate the UV behaviour of the amplitude, leading to a celestial amplitude which is a meromorphic function of $\Delta+\Delta'$.

In particular, the celestial eikonal-probe amplitude on the Schwarzschild background is given by the Mellin transform:
\begin{equation}
    \begin{split}
        \widetilde{M}^{\mathrm{Schw}}_{2}  =& \int_0^{\infty} \d\omega \int_0^{\infty} \d\omega' \, \omega^{\Delta-1}\,(\omega')^{\Delta'-1}\,\frac{\pi\,\delta(p_0'-p_0)}{2\,M}\, \cM_{\mathrm{eik}}^{\mathrm{Schw}} \\
        =& \frac{8\pi^2\,G\,M}{f(z,z')}\,\frac{(1+|z|^2)^{\Delta'+1}}{(1+|z'|^2)^{\Delta'}}\int_0^\infty \d\omega\, \omega^{\Delta+\Delta'-2} \, \exp\!\left(2\mathrm{i}\,\gamma_E\,G\,M\,(1+|z|^2)\, \omega\right) \\
        & \times\,\exp\!\left[ \sum_{k\geq1} \frac{ \, 2\,\zeta(2k+1)}{2k+1}  \left(\mathrm{i} \, G\, M\,(1+|z|^2)\, \omega\right)^{2k+1}\right] \\
        &\times\,\exp\!\left[-\mathrm{i}\, G\, M\, (1+|z_1|^2)\,\omega\,\log\!\left(\frac{4\,\mu^2}{\omega^2 \,f(z,z')} \, \frac{1+|z'|^2}{1+|z|^2} \right) \right] \, ,
    \end{split}
\end{equation}
where
\be\label{angfunct}
f(z,z'):=\left|z-\frac{1+|z|^2}{1+|z'|^2}\,z'\right|^2\,,
\ee
is a function of the two points on the celestial sphere\footnote{Note that all dependence on the scales $G$ and $M$ can be scaled out of the amplitude as an overall factor, although we will not do this explicitly for the black hole eikonal-probe formulae.}. The remaining integral is in the same class \eqref{Mellinmaster} that we have encountered before, so the celestial amplitude exists as a meromorphic function of $\Delta+\Delta'$.

The celestial eikonal-probe amplitude on the Schwarzschild metric can also be approximated as an asymptotic series:
\begin{multline}\label{cepSchw}
\widetilde{M}_{2}^{\mathrm{Schw}} =\frac{8\pi^2\,G\,M}{f(z,z')}\,\frac{(1+|z|^2)^{\Delta'+1}}{(1+|z'|^2)^{\Delta'}}\, \exp\!\left[2\mathrm{i}\, G\, M\,(1+|z|^2)\,\e^{\partial_{\Delta+\Delta'}}\, \partial_{\Delta+\Delta'}\right]\\
\times\,  \exp\!\left[ \sum_{k\geq1} \frac{ 2\,\zeta(2k+1)}{2k+1}\,\left(\mathrm{i} G\, M\,(1+|z|^2)\, \e^{\partial_{\Delta+\Delta'}} \right)^{2k+1}\right]\, \mathcal{F}(\Delta+\Delta')_{\text{Schw}} \, ,
\end{multline}
where the `primary' function $\mathcal{F}_{\mathrm{Schw}}$ is analogous to the function \eqref{eq:calF2} appearing in the shockwave calculation, albeit with slightly more complicated kinematic dependence:
\begin{equation}\label{Sch_BH_primary}
\mathcal{F}(\Delta+\Delta')_{\text{Schw}}= \frac{\mathrm{i}^{\Delta+\Delta'-1}\,\Gamma(\Delta+\Delta'-1)}{\left[G\,M\,(1+|z|^2)\,\log\!\left(\frac{C_E\,f(z,z')}{4\,\mu^2}\,\frac{1+|z|^2}{1+|z'|^2}\right)\right]^{\Delta+\Delta'-1}} \, .
\end{equation}
It is easy to see that this amplitude has a divergent Born approximation (as expected) and exhibits the \emph{same} poles in the $(\Delta+\Delta')$-plane as the shockwave amplitude: namely, it has poles along the non-positive real axis of the form \eqref{delres1}. Finally, the behaviour of $\widetilde{M}^{\mathrm{Schw}}_{2}$ for large values of $\mathrm{Re}(\Delta+\Delta')$ is practically identical to \eqref{epsadd3}. 


\subsubsection{Kerr}

The evaluation of the celestial eikonal-probe amplitude for a massless scalar on a Kerr black hole metric proceeds along similar lines. The starting point is the momentum space eikonal probe amplitude \eqref{eikprobeKerr}, which differs from the Schwarzschild amplitude only by a spin-dependent phase:
\be
\cM_{\mathrm{eik}}^{\mathrm{Kerr}}=\exp\!\left[\im\,a\,(\omega'\,|z'|^2-\omega\,|z|^2)\right]\cM_{\mathrm{eik}}^{\mathrm{Schw}}\,,
\ee
where, without loss of generality, we have chosen the spin of the Kerr metric orthogonal to the $xy$-plane and $a$ denotes the Kerr parameter. 

Although the Kerr and Schwarzschild amplitudes only differ by a phase in momentum space, this phase depends on the frequencies of the incoming and outgoing probe particles. This has interesting implications for the Mellin transform. To begin with, the Born amplitude associated with the Kerr metric,
\be\label{BornKerr}
\cM_{\mathrm{Born}}^{\mathrm{Kerr}}=\frac{8\pi\,G\,\alpha(s)^2}{t}\,\exp\!\left[\im\,a\,(\omega'\,|z'|^2-\omega\,|z|^2)\right]\,,
\ee
has a \emph{finite} Mellin transform, as the spin-dependent phase now regulates the high-frequency region of the integrals\footnote{Although the eikonal-probe amplitude, and the Born amplitude in particular, is sensitive only to linear in $G$ terms in a large-distance expansion of the metric~\cite{Adamo:2021rfq}, this still contains multipole moments of \emph{all} orders~\cite{Vines:2017hyw}, effectively meaning that the amplitudes are sensitive to the finite-size of the source of the Kerr metric~\cite{Israel:1970kp,Israel:1976vc,Balasin:1993kf}.}~\cite{Gonzo:2022tjm}. Therefore, in the case of the Kerr metric, the eikonal phase is not needed to obtain a finite result, although it \emph{does} lead to an all-orders in $G$ completion of the Born amplitude.

The asymptotic series representation of the celestial eikonal-probe amplitude on the Kerr background is given by:
\begin{multline}\label{cepKerr}
\widetilde{M}_{2}^{\mathrm{Kerr}} =\frac{8\pi^2\,G\,M}{f(z,z')}\,\frac{(1+|z|^2)^{\Delta'+1}}{(1+|z'|^2)^{\Delta'}}\, \exp\!\left[2\mathrm{i}\, G\, M\,(1+|z|^2)\,\e^{\partial_{\Delta+\Delta'}}\, \partial_{\Delta+\Delta'}\right]\\
\times\,  \exp\!\left[ \sum_{k\geq1} \frac{ 2\,\zeta(2k+1)}{2k+1}\,\left(\mathrm{i} G\, M\,(1+|z|^2)\, \e^{\partial_{\Delta+\Delta'}} \right)^{2k+1}\right]\, \mathcal{F}(\Delta+\Delta')_{\text{Kerr}} \, ,
\end{multline}
where
\begin{equation}\label{Kerr_BH_primary}
\mathcal{F}(\Delta+\Delta')_{\text{Kerr}}= \frac{\mathrm{i}^{\Delta+\Delta'-1}\,\Gamma(\Delta+\Delta'-1)}{\left[G\,M\,(1+|z|^2)\,\log\!\left(\frac{C_E\,f(z,z')}{4\,\mu^2}\,\frac{1+|z|^2}{1+|z'|^2}\right)+a\,\frac{(|z|^2-|z'|^2)}{1+|z'|^2}\right]^{\Delta+\Delta'-1}} \, .
\end{equation}
Observe that, due to its dependence on the Kerr parameter $a$, the function $\mathcal{F}(\Delta+\Delta')_{\mathrm{Kerr}}$ has a finite $G\to0$ limit, corresponding to the underlying finite celestial Born amplitude. In all other respects, the analytic and asymptotic properties of the Kerr eikonal-probe amplitude are identical to those of the Schwarzschild amplitude \eqref{cepSchw}, although the residues of the poles of \eqref{cepKerr} will depend on the Kerr parameter $a$ as well as the probe kinematics.


\subsection{Eikonal to eikonal-probe matching}

The structure of the celestial eikonal amplitude \eqref{Aeikfact} clearly shares many similarities with the shockwave celestial eikonal probe amplitude \eqref{eq:tildeM2}. In momentum space, there is an explicit matching between the 4-point eikonal amplitude of gravitationally-coupled massless scalars and the 2-point eikonal-probe amplitude on the shockwave background, so one naturally expects this correspondence to persist in some guise for the celestial amplitudes. 

However, there are several features of the celestial amplitudes which mean that matching between constituent pieces of the eikonal and eikonal-probe amplitudes must be taken with care. The fact that the Mellin transform on both sides of this correspondence is taken with respect to an energy scale means that matching between the celestial amplitudes involves some \emph{a priori} strange-looking replacements, including some which change the physical dimensions of a given quantity. This is also a consequence of the fact that $\cA_{\mathrm{eik}}$ has no energy scales, depending instead on the four conformal dimensions of the external scalars through $\beta$, while $\widetilde{M}^{\mathrm{shock}}_2$ depends on the lightfront energy of the shockwave background, $P_-$. Furthermore, the correspondence \eqref{eikprobeUR} between eikonal and eikonal-probe amplitudes in momentum space involves an energy-conserving delta function whose interplay with the Mellin transform is non-trivial.

Let $\rho$ denote the replacement map which acts on Newton's constant, $\beta$ and the conformal cross-ratio $z$ as:
\begin{equation}\label{rhomap1}
    \begin{split}
        &G\rightarrow G\,P_-  \, ,\\
        &z\rightarrow |z-z'|^2 \, ,\\
        &\beta \rightarrow 2\,(\Delta+\Delta'-2) \, .
    \end{split}
\end{equation} 
It is easy to verify, by comparison with \eqref{eq:Abeta_eik} and \eqref{2pt_function} that this replacement map acts (formally) on the 4-point celestial eikonal amplitude as:
\be\label{rhomap2}
\rho\!\left(\cA_{\mathrm{eik}}(\beta,z)\right)=-\frac{1}{\pi^2}\,\exp\!\left[\im\,\e^{\partial_{\Delta+\Delta'}}\,\partial_{\Delta+\Delta'}\right]\,\widetilde{M}_2^{\mathrm{shock}}(\Delta+\Delta',z,z')\,.
\ee
In particular, the celestial eikonal amplitude is mapped to the celestial eikonal-probe amplitude up to an overall shift operator and numerical factor. 

Of course, \eqref{rhomap2} should be interpreted with care, as it essentially makes sense only at the level of the asymptotic series approximations of the celestial amplitudes on both sides. In any case, observe that while the overarching correspondence between $\cA_{\mathrm{eik}}$ and $\widetilde{M}_2^{\mathrm{shock}}$ agrees with that predicted by~\cite{deGioia:2022fcn}, the precise details of the replacement map are new.




\section{Dispersion and monodromy relations}\label{sec:dispersion}

Dispersion relations are powerful non-perturbative constraints on analytically continued scattering amplitudes with complex kinematics, following from basic assumptions such as analyticity and causality of S-matrices in 4d Minkowski space. Cauchy's residue theorem allows for dispersion relations relating real and imaginary parts of amplitudes. In the context of celestial amplitudes, one can select appropriate contours in the complex plane of energy scales to turn the analytically-continued Mellin transform into a dispersion relation constraining the celestial amplitude in terms of poles in the energy spectrum. 

As they are all-orders in the coupling (in the sense that we have integrated out the dependence in energy scales), celestial eikonal and eikonal-probe amplitudes are ideal candidates for such a dispersion analysis. Although we were only able to formally write the Mellin integrals as exponentiated derivative operators, we shall see that it is possible to write down combinations of the real and imaginary parts of the integrals just using analyticity properties in the complex plane. In this section we obtain a dispersion relation for $\cA_{\mathrm{eik}}$, the dynamical part of the 4-point celestial eikonal amplitude for gravitational scattering of massless scalars, as well as a monodromy relation for $\widetilde{M}_2^{\mathrm{shock}}$, the celestial eikonal-probe amplitude for a massless scalar scattering on a gravitational shockwave. Note that dispersion relations of celestial amplitudes have been studied before~\cite{Chang:2021wvv,Garcia-Sepulveda:2022lga,Ghosh:2022net} in different contexts.


\subsection{Dispersion relation for celestial 4-point eikonal amplitudes}

\renewcommand{\o}{\omega}

To find a dispersion or monodromy relation between the real and imaginary parts of the celestial 4-point eikonal amplitude \eqref{Aeikfact}, we go back to the Mellin integral:
\begin{equation}\label{4ptdis1}
\cA_{\mathrm{eik}}(\beta,z)=\frac{-1}{2\,G^{\beta/2}\,z}\int_{0}^{\infty}\d\omega\,\omega^{\beta/2}\frac{\Gamma(-\mathrm{i}\,\omega)}{\Gamma(\mathrm{i}\,\omega)} \left( \frac{4\, \tilde\mu^2}{\omega\, z}\right)^{-\mathrm{i}\, \omega}\,.
\end{equation}
The integrand here has a sequence of poles in the complex $\omega$-plane located at
\begin{equation}
    \omega = \e^{\mathrm{i}\frac{3\pi}{2}}\,n = -\mathrm{i}\,n\,,\qquad n\in\mathbb{Z}_{>0}\,,
\end{equation}
with a branch cut running along the positive real half-axis. 

\begin{figure}[htp]
    \centering
    \begin{tikzpicture}[font=\Large]
    \node[cross=4pt, black, label={[black]above right:$\omega$}] at (2,1) {};
    \node[cross=4pt, red, label={[red]below:$C_1$}] at (1.6,-0.25) {};
    \node[cross=4pt, red, label={[red]below:$C_2$}] at (-1.6,-0.25) {};
    \draw[thick] (-3,0) -- (0,0);
    \draw[thick,->] (0,0) -- (0,1);
    \cross{0,-0.7};
    \cross{0,-1.4};
    \cross{0,-2.1};
    \cross{0,-2.8};
    \draw[red,thick]{(3.2,-0.2)--(0.2,-0.2)};
    \draw[red,thick]{(-3.2,-0.2)--(-0.2,-0.2)};
    \draw[red, thick] (-3.2,-0.2) arc (0:180:-3.2) node[midway, below] {$C_R$};
    \draw[red, thick] (-0.2,-0.2) arc (0:180:-0.2) node[midway,below right] {$C_{\epsilon}$};
    \draw[thick] (0,-0.1)
    \foreach \x in {0.3,0.6,...,1.8} {
      -- ++(0.3,0.2) -- ++(0.3,-0.2)
    }
    -- (3,-0.1);
\end{tikzpicture}
    \caption{Contour integral of celestial 4-point eikonal amplitude in the complex $\omega$ plane.}
    \label{fig:4pt}
\end{figure}
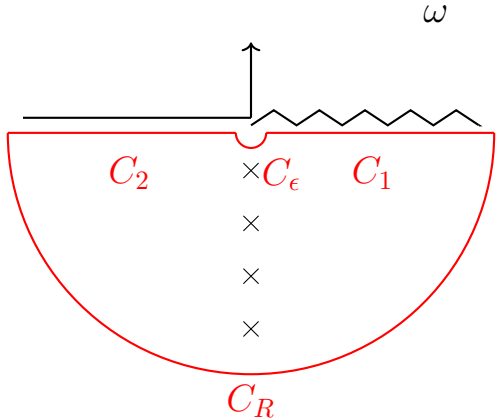

The Mellin transform in \eqref{4ptdis1} can now be analytically continued to a clockwise contour in the complex $s$-plane as illustrated in Figure~\ref{fig:4pt}, with the integral decomposing into four distinct pieces
\begin{equation}
\begin{split}
    I_{C} 
    &=\frac{-1}{2\,G^{\beta/2}\,z} \oint_C \d\omega \, \omega^{\frac{\beta}{2}}\,\frac{\Gamma(-\mathrm{i}\,\omega)}{\Gamma(\mathrm{i}\,\omega)} \left( \frac{4\,\tilde\mu^2}{\omega \,z}\right)^{-\im\,\omega} \\
    &= I_1+I_2 +I_{\epsilon}+I_R \, ,
\end{split}
\end{equation}
where the decomposition of the contour $C=C_1\cup C_R\cup C_2\cup C_\epsilon$ corresponds to each of the four integrals in the second line, with $C_R$ the semi-circle of radius $R$ and $C_\epsilon$ the semi-circle of radius $\epsilon$.

It can be shown that in the limits where $\epsilon\to0$ and $R\to\infty$, the contributions from $I_\epsilon$ and $I_R$ both vanish. Therefore, by Cauchy's theorem we have
\begin{equation}
    I_1 +I_2 = 2\pi \mathrm{i} \, \sum_{n\geq1}\text{Res}\!\left[\frac{-1}{2\,G^{\beta/2}\,z} \, \omega^{\frac{\beta}{2}}\,\frac{\Gamma(-\mathrm{i}\,\omega)}{\Gamma(\mathrm{i}\,\omega)} \left( \frac{4\,\tilde\mu^2}{\omega \,z}\right)^{-\mathrm{i}\,\omega},\:\omega=-\im\,n\right] \, .  \label{eq:dispfromcomplexs}
\end{equation}
The integrals $I_1$ and $I_2$ are related to the celestial 4-point eikonal amplitudes as
\begin{equation}
    I_1 =\frac{-1}{2\,G^{\beta/2}\,z}\int_{\infty}^0 \d x \, (x\, \e^{2\pi\mathrm{i}})^{\frac{\beta}{2}}\,\frac{\Gamma(-\mathrm{i}\,x)}{\Gamma(\mathrm{i}\,x)}\left(\frac{4\,\tilde\mu^2}{x\, z}\right)^{-\mathrm{i}\,x} =-\e^{\mathrm{i}\pi\, \beta}\,\mathcal{A}_{\text{eik}}(\beta,z) \, .
\end{equation}
\begin{equation}
        I_2 = \frac{-1}{2\,G^{\beta/2}\,z}\int_0^{\infty} \d x \, \e^{\mathrm{i}\pi} (x\, \e^{\mathrm{i}\pi})^{\frac{\beta}{2}} \,  \frac{\Gamma(\mathrm{i}\,x)}{\Gamma(-\mathrm{i}\,x)}\left(\frac{4\,\tilde\mu^2}{-x\,z}\right)^{\mathrm{i}\,x} =-\e^{\frac{\mathrm{i} \pi\, \beta}{2}}\, \overline{\mathcal{A}_{\text{eik}}}(\bar{\beta},-z) \, ,
\end{equation}
where $\overline{\mathcal{A}_{\text{eik}}}(\bar{\beta},-z)$ denotes the complex conjugate of \eqref{eq:Abeta_eik} evaluated on $\bar{\beta}$ and $-z$.

Evaluating the residues on the right-hand side of \eqref{eq:dispfromcomplexs} now gives the dispersion relation
\begin{equation}
\mathcal{A}_{\text{eik}}(\beta,z)+\e^{-\frac{\mathrm{i} \pi \beta}{2}}\, \overline{\mathcal{A}_{\text{eik}}}(\bar{\beta},-z) = -\frac{\pi}{z}\,\e^{-\im\pi\beta} \sum_{n\geq1}\frac{\mathrm{i}^n}{n!\, (n-1)!}\left(\frac{-\mathrm{i}\,n}{G}\right)^{\frac{\beta}{2}}\left(\frac{n\, z}{4\, \tilde{\mu}^2}\right)^n \, . \label{eq:disperAandA*}
\end{equation}
This relates the real and imaginary parts of the 4-point eikonal celestial amplitude in terms of a formal sum of residues\footnote{The sum on the right-hand-side of the dispersion relation \eqref{eq:disperAandA*} is difficult to evaluate because each term behaves roughly like $n^n$. Using Stirling's approximation, one can trade $n^n$ for a factorial which makes the sum tractable. Adding more terms from Stirling's approximation surprisingly results in better and better approximations for the series, although the asymptotic nature of the approximation means that these improvements eventually cease.}. 

At fixed $\beta$, it is clear that this sum has an infinite radius of convergence in $z$ and resembles  $e^{i c z}(z^{(\beta-1)/2}+O(z^{(\beta-3)/2}))$ where $c\sim e/(4\tilde \mu^2)$.


\subsection{Monodromy relation for celestial eikonal-probe amplitudes}

A similar analysis can be performed for celestial eikonal-probe amplitudes, by analytically continuing the Mellin integral in each case to a certain closed contour in the complex $\omega$-plane. We will carry this out for the specific example of the eikonal-probe amplitude on the gravitational shockwave background, but the basic argument also works for the Schwarzschild and Kerr cases.

Consider the analytically continued contour integral
\begin{equation}
\begin{split}
     I_{C} &= \frac{2\pi^2}{(GP_{-})^{\Delta+\Delta'-2}\,|z-z'|^2} \oint_C \d\omega \, \omega^{\Delta+\Delta'-2}\, \frac{\Gamma(-\mathrm{i}\,\omega)}{\Gamma(\mathrm{i}\,\omega)} \left(\frac{4\, \tilde\mu^2}{\omega^2\,|z-z'|^2}\right)^{-\mathrm{i}\, \omega} \\
     &=I_1+I_2+I_R+I_{\epsilon} \, \label{eq:IC2}
\end{split}
\end{equation}
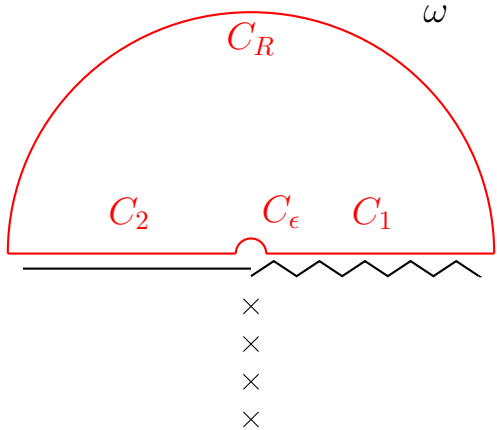
\begin{figure}[htp]
    \centering
    \begin{tikzpicture}[font=\Large]
    \node[cross=4pt, black, label={[black]above right:$\omega$}] at (2,3) {};
    \node[cross=4pt, red, label={[red]above:$C_1$}] at (1.6,0.25) {};
    \node[cross=4pt, red, label={[red]above:$C_2$}] at (-1.6,0.25) {};
    \draw[thick] (-3,0) -- (0,0);
    \cross{0,-0.5};
    \cross{0,-1};
    \cross{0,-1.5};
    \cross{0,-2};
    \draw[red,thick]{(3.2,0.2)--(0.2,0.2)};
    \draw[red,thick]{(-3.2,0.2)--(-0.2,0.2)};
    \draw[red, thick] (3.2,0.2) arc (0:180:3.2) node[midway, below] {$C_R$};
    \draw[red, thick] (0.2,0.2) arc (0:180:0.2) node[midway,above right] {$C_{\epsilon}$};
    \draw[thick] (0,-0.1)
    \foreach \x in {0.3,0.6,...,1.8} {
      -- ++(0.3,0.2) -- ++(0.3,-0.2)
    }
    -- (3,-0.1);
\end{tikzpicture}
    \caption{Contour of celestial eikonal-probe amplitude in the shockwave background.}
    \label{fig:contourof2pt}
\end{figure}
where the anti-clockwise contour in the upper half-plane is illustrated in Figure~\ref{fig:contourof2pt}. The branch cut of the integrand has been chosen from $0$ to $+\infty$ on the real axis, so that a complex $\omega= x\,\e^{\mathrm{i}\theta}$ for $\theta \in (0,2\pi)$. The poles of the integrand in \eqref{eq:IC2} are located at
\begin{equation}
    \omega = \e^{\mathrm{i}\frac{3\pi}{2}}\, n \, , \qquad n\in\Z_{>0} \, ,
\end{equation}
and these are \emph{not} enclosed by the contour $C$.

In the limits where $\epsilon\to0$ and $R\to\infty$, the contributions from $I_{\epsilon}$ and $I_R$ can be shown to vanish, so that Cauchy's theorem gives
\begin{equation}
I_1+I_2 = 0 \, .
\end{equation}
The remaining integrals can then be shown to obey
\begin{equation}
    I_1 = \widetilde{M}^{\mathrm{shock}}_2(\Delta+\Delta',\, z,z') \, ,
\end{equation}
and 
\begin{equation}
    I_2 = -\e^{\mathrm{i}\pi(\Delta+\Delta')}\,\overline{\widetilde{M}_2^{\mathrm{shock}}}(\bar{\Delta}+\overline{\Delta'},\, z,z') \, ,
\end{equation}
leading to the monodromy relation,
\begin{equation}\label{monrel}
    \widetilde{M}^{\mathrm{shock}}_2(\Delta+\Delta',\, z,z') = \e^{\mathrm{i}\pi(\Delta+\Delta')}\,\overline{\widetilde{M}_2^{\mathrm{shock}}}(\bar{\Delta}+\overline{\Delta'},\, z,z') \, ,
\end{equation}
for the celestial eikonal-probe amplitude.

Note that one might have hoped to complete this to a dispersion relation by using a contour in the lower half-plane, instead of the contour in Figure~\ref{fig:contourof2pt}. However, the integral over the large radius half-circle in the lower half-plane is non-vanishing for the eikonal-probe amplitude (in contrast to the 4-point eikonal amplitude). It would be interesting to explore whether it is possible to get such a relation by some other means. 

Finally, we observe that the celestial eikonal-probe amplitudes for the Schwarzschild and Kerr spacetimes obey precisely the same monodromy relation \eqref{monrel} as the shockwave. Consequently, one could hope that there are corresponding dispersion relations for 4-point celestial amplitudes involving two massless and two \emph{massive} legs. Unfortunately, the hyperbolic integrals needed to define the conformal primary states for the massive legs~\cite{Pasterski:2017kqt} are significantly more complicated than a Mellin transform, and we have not yet managed any explicit calculations in this regard.


\section{Carrollian eikonal probe amplitudes}\label{sec:carrollian_probe}

In contrast to celestial amplitudes, \emph{Carrollian amplitudes} of massless particles can be defined as position-space scattering amplitudes with external states parametrized by an insertion location on $\scri$, the null conformal boundary of an asymptotically flat spacetime~\cite{Bagchi:2022emh,Donnay:2022aba,Donnay:2022wvx,Mason:2023mti,Salzer:2023jqv,Nguyen:2023miw,Bagchi:2023cen,Liu:2024nfc,Have:2024dff,Stieberger:2024shv}. This conformal boundary has two distinct components, $\scri=\scri^-\cup\scri^+$ corresponding to the past/future null infinities, each of which has the topology of a lightcone: $\scri^{\pm}\cong\R\times S^2$. Consequently, the external states of an `all-outgoing' Carrollian amplitude are parametrized by points $(u,z,\bar{z})\in\scri^+$, where $u$ is the retarded Bondi time coordinate along the null generators of $\scri^+$.

As $u$ can be regarded as the canonical conjugate variable to the frequency of a null 4-momentum, in the first instance Carrollian amplitudes can be defined from momentum space amplitudes through a Fourier transform in the frequency of the external states:
\begin{equation}
        C_n(\{u_i, z_i,\bar{z}_i\}) =\int_{0}^{\infty} \prod_{i=1}^n \frac{\d\omega_i}{2\pi}\, \e^{\im\epsilon_i\,\omega_i\, u_i}\,\cM_{n}(\{\omega_i,z_i,\bar{z}_i\}) \, ,  \label{eq:nCarrollian}
\end{equation}
where $\epsilon_i=\pm$ determines whether the states is outgoing or incoming. Due to the momentum conserving delta functions appearing in the momentum space amplitude, the resulting Carrollian amplitude is generically divergent. For instance, the two-point Carrollian amplitude for massless scalars in Minkowski spacetime can be written in a regulated form as~\cite{Donnay:2022wvx}
\begin{multline}
    C_2(u_1, z_1,\bar{z}_1; u_2, z_2,\bar{z}_2) \\
    = \frac{1}{4\pi}\,\lim_{\varepsilon\rightarrow0}\left[ \frac{1}{\varepsilon}-\left(\gamma_E+\ln|u_{12}| +\frac{\im\,\pi}{2}\,\text{sign}(u_{12})\right)\right]\,\delta(z_{12})\,\delta(\bar{z}_{12}) \, ,
\end{multline}
for $u_{12}:=u_1-u_2$, $z_{12}:=z_1-z_2$, etc. Clearly, this result is both divergent and distributional, containing delta functions in the locations on the celestial sphere. The divergence is typically dealt with by taking $u$-derivatives to obtain a two-point function which is of the `electric' type for a Carrollian CFT~\cite{Donnay:2022wvx,Mason:2023mti}; from the bulk side, this corresponds to scattering the characteristic data of the linear fields rather than the fields themselves~\cite{Penrose:1980yx}.

However, we observe that both the divergent and distributional properties of the 2-point Carrollian amplitude are cured when one instead considers Carrollian eikonal-probe scattering on a curved spacetime. To see this, we use the `modified Mellin transform' of Banerjee~\cite{Banerjee:2018gce,Banerjee:2019prz,Banerjee:2020kaa}, which combines the Fourier and Mellin transforms to define a modified Mellin amplitude:
\be\label{eq:modMellin}
    \widetilde{\mathcal{M}}_{n,\text{mod}}(\{u_i, z_i,\bar{z}_i,\Delta_i\})=\int_{0}^{\infty} \prod_{i=1}^n \frac{\d\omega_i}{2\pi}\, \e^{\im\epsilon_i\, \omega_i\, u_i} \, \omega_i^{\Delta_i-1}\,\mathcal{M}_n(\{\omega_i, z_i,\bar{z}_i\}) \, . 
\ee
Comparing \eqref{eq:nCarrollian} with \eqref{eq:modMellin}, one immediately sees that the Carrollian amplitudes are obtained from modified Mellin amplitudes by setting all conformal dimensions to 1: 
\be\label{modMelmatch}
    C_n(\{u_i, z_i,\bar{z}_i\})= \left.\widetilde{\mathcal{M}}_{n,\text{mod}}(\{u_i, z_i,\bar{z}_i,\Delta_i\})\right|_{\Delta_1=\cdots=\Delta_n=1}\,.
\ee
The upshot of this fact is that all of our results on the analyticity of celestial eikonal-probe amplitudes can be trivially extended to Carrollian eikonal-probe amplitudes.

Consider, for example, the case of eikonal-probe scattering on the gravitational shockwave. Based on \eqref{modMelmatch}, the Carrollian amplitude is given by
\be\label{Carep1}
C_2^{\mathrm{shock}}=\left.\frac{-2\pi^2\,G\,P_-}{|z_{12}|^2} \int_0^{\infty} \d\omega\, \omega^{\Delta-2}\,\frac{\Gamma(-\im\,G\,P_-\,\omega)}{\Gamma(\im\,G\,P_-\,\omega)}\left(\frac{4\,\tilde{\mu}^2}{\omega^2\,|z_{12}|^2}\right)^{-\im G\,P_-\,\omega}\,\e^{\im\,\omega\,u_{12}}\right|_{\Delta=2}\,.
\ee
The only difference from the celestial case that we analyzed above is the presence of $\e^{\im\omega u_{12}}$ in the integrand, but this can simply be absorbed into the eikonal phase itself and does not affect the analyticity properties of the integral. In particular, as the integral is analytic at $\Delta=2$, reduction from the modified Mellin amplitude to the Carrollian amplitude is straightforward and the result will be an analytic function of $u_{12}$ as well as the coordinates on the celestial sphere.

The Carrollian eikonal-probe amplitude also admits an asymptotic series approximation in terms of exponentiated differential operators. Defining
\begin{equation}
    \mathcal{F}(\Delta, u_{21})
    :=\frac{\Gamma(\Delta-1)}{4\,\pi^2}\,\frac{\mathrm{i}^{\Delta-1}}{\left[u_{21}+G\, P_- \, \log\!\left(\frac{C_E \, |z_{12}|^2}{4\,\mu^2}\right)\right]^{\Delta-1}} \, ,
\end{equation}
this asymptotic series is given by
\begin{multline}\label{2pt_functionmod}
    C_{2}^{\text{shock}} =\frac{2\pi^2\, P_- \,G}{|z_{12}|^2}\,
    \exp\!\left(2\mathrm{i}\, G\,  P_-\, \partial_{\Delta} \, \e^{\partial_{\Delta}} \right) \\
    \times\,\left.\exp\!\left[ \sum_{k\geq1} \frac{ 2\,\zeta(2k+1)}{2k+1}  (\mathrm{i} \, G\,  \, P_-\, \e^{\partial_{\Delta}} )^{2k+1}\right] \mathcal{F}(\Delta, u_{21})\,\right\vert_{\Delta=2} \, .
\end{multline}
A similar story also holds for the Carrollian eikonal-probe amplitudes on black hole spacetimes, of course.


\section{Discussion}
\label{sec:discussion}

In this paper, we have seen that taking eikonal exponentiation into account leads to well-defined gravitational celestial amplitudes, in the sense that the resulting expressions are meromorphic functions of the conformal dimensions with isolated singularities that can be precisely characterised. The resulting celestial eikonal and eikonal-probe amplitudes have many remarkable properties, and we believe that there are many interesting avenues for future research based on these findings.

We conclude with a brief discussion of effects beyond the leading eikonal approximation, before touching on a list of potential topics for further study.


\subsection{Beyond leading eikonal: absorptive effects} \label{sec:subleading_eikonal}

\paragraph{Sub-leading eikonal corrections:} It is possible to consider sub-leading corrections to the eikonal approximation, corresponding to the appearance of loop, rather than exclusively tree-level, diagrams in the resummation of the perturbative series. This was first studied by Amati-Ciafaloni-Veneziano (ACV)~\cite{Amati:1988tn,Amati:1990xe} in the case of massless scalar scattering, who showed that the next-to-leading order (NLO) and next-to-next-to-leading order (NNLO) corrections to the leading eikonal amplitudes can also be exponentially resummed. A nice modern take on the summary of the results in various dimensions can be found in~\cite{Haring:2022cyf}.

This is neatly captured by exponentiated phases, written in terms of the Mandelstam variables \eqref{shockMand} and impact parameter $|x^{\perp}|$:
\begin{equation}
\begin{split}
     \delta_0 &= -\frac{G\,s}{\hbar}\, \log (\mu\, |x^\perp|) \,, \qquad \delta_1 = \frac{6 }{\pi}\,\frac{G^2\, s}{|x^\perp|^2}\, \log s \, , \\
     \delta_2 & = \frac{2 G^3\, s^2}{\hbar \,|x^\perp|^2}+\mathrm{i}\underbrace{\frac{2 G^3\, s^2}{\hbar \,|x^\perp|^2}\, \frac{\log s}{\pi}\left(\log \frac{|x^\perp|^2}{\mu^2} +2\right)}_{\text{Imaginary part}} \, . \label{eq:ACV}
\end{split}
\end{equation}
Here, $\delta_0$ is the leading eikonal phase, while $\delta_1$ and $\delta_2$ are the NLO and NNLO corrections, respectively. We have temporarily restored explicit $\hbar$-dependence in these expressions to illustrate the following important point: $\delta_0$ and $\delta_2$ go like $\hbar^{-1}$, while $\delta_1$ goes like $\hbar^0$. Consequently, $\delta_0$ and $\delta_2$ represent the dominant \emph{classical} contributions to the amplitude, while $\delta_1$ is quantum in nature\footnote{As such, it is not necessary to actually exponentiate $\delta_1$, as this just corresponds to a rearrangement of quantum contributions to the amplitude. By contrast, exponentiation of $\delta_0$ is essential for the amplitude to have a well-defined classical limit.}. This is entirely consistent with the non-zero NLO contribution to the \emph{massive} case~\cite{KoemansCollado:2019ggb,Adamo:2022ooq}, whose ultra-relativistic limit vanishes.

Due to the imaginary part of $\delta_2$, the NNLO correction to the eikonal phase leads to an exponential damping factor in the eikonal-probe amplitude. This damping behaviour is absorptive, and can be seen as the first hint of the amplitude resolving some finite-size features of the background. ACV also famously showed that this absorptive effect has an essentially \emph{radiative} nature: the imaginary part of the NNLO phase is due to the exclusion of soft gravitational radiation in the outgoing state~\cite{Amati:1990xe}. Indeed, this imaginary part disappears when one considers an operator that also includes emission~\cite{DiVecchia:2022nna}. In other words, as one relaxes the strict eikonal limit, a $2\to2$ process (in flat space) or $1\to1$ probe scattering (in a background) without any bremsstrahlung becomes less likely, meaning that the amplitudes for such non-radiative scattering processes become suppressed.

Explicitly performing the impact parameter integral for the eikonal amplitude with $\delta_2$ has not yet been achieved\footnote{Even for the inclusion of $\delta_1$, the eikonal integrals were only evaluated explicitly for general mass configurations recently~\cite{Adamo:2022ooq}.}, but it can be approximated by using the leading-order saddle point:
\begin{equation}
     |x^{\perp}|_{*}^2= \frac{4 G^2\,s^2}{-t} = \frac{4 G^2  \, \omega^2}{z} \, ,
\end{equation}
having made use of the kinematics of \eqref{AAmp}. Evaluated on this saddle point, one finds that 
\begin{equation} 
     \mathrm{Im}\,\delta_2\Big|_{|x^{\perp}|_{*}} 
     =C\,\omega^2\,\left(\log \omega\right)^2 +O(\omega^2\,\log\omega) \, ,
\end{equation}
where $C$ is a positive coefficient that does not depends on $\omega$. Thus, the inclusion of $\delta_2$ introduces an exponential damping factor of the form $\e^{-\omega^2\,(\log\omega)^2}$ into the Mellin transform (ignoring all overall constants). Note that this damping factor decays faster than the $\e^{-\omega^2}$ due to black hole production discussed in~\cite{Arkani-Hamed:2020gyp}, although they also come with different powers of $G$.  

Let us make one somewhat speculative remark. The convergence of the Mellin transform defining celestial eikonal amplitudes relied on the high-energy damping of the gravitational Born approximation by the eikonal phase. Our arguments for the existence of this Mellin integral was based on the analytic continuation of the eikonal Gamma function \eqref{Mellinmaster}; as reviewed in Appendix~\ref{appendix:eik_gamma}, when rotating the contour of integration for the eikonal Gamma function to the imaginary axis, it is necessary to include an $\im\epsilon$-regulator to ensure convergence of the analytic continuation. The dependence on $\epsilon$ simply drops out after this integral has been performed, making its appearance an essentially formal mathematical artefact. Remarkably, the NNLO eikonal phase seems to provide a \emph{physical} origin for this damping: it can be viewed as a sort of $\im\epsilon$-regulator which emerges organically from the inclusion of inelastic effects in trans-Planckian scattering.

\paragraph{Stringy corrections:} ACV also studied corrections to the leading eikonal approximation coming from string effects~\cite{Amati:1987uf,Amati:1988tn}. These are not \textit{per se} fixed angles effects, but rather fixed \emph{impact parameter} effects, so their interpretation on the celestial sphere is not entirely obvious. Indeed, doing the Mellin transform at fixed $z$ for $s=\omega^2$ and $t=-z \omega^2$ interacts non-trivially with the transform from impact parameter space.

Given this word of caution, let us briefly analyze briefly their effects on the celestial amplitudes. The stringy effects identified by ACV come in two basic types: `diffractive' effects due to the production of energetic two-body scattering, and `inelastic' effects coming from the production of very massive strings or a large number of strings. In $d$ spacetime dimensions, these lead to a modification of the celestial amplitude of the form (cf., equation 6.34 of~\cite{Amati:1987uf})
\begin{align}
    \cM_{\rm dif.}(s,t)&=\cM_{\rm eik}(s,t)\,\exp\!\left[-c\, \frac{\sqrt{s}}{R_S}\,\theta^{(d-2)(d-3)}\right]\\
    \cM_{\rm inel.}(s,t)&=\cM_{\rm eik}(s,t)\,\exp\!\left[-g_s\,\frac{s^2}{(\log(s))^{d/2-1}}\right]\,,
\end{align}
where $R_S= (2 G \sqrt{s})^{1/(d-3)}$ is the Schwarzschild radius, $\cos\theta=1+2t/s$ is the (cosine of the) scattering angle and $g_s$ is the string coupling.

Interestingly, in $d=4$, the Schwarzschild radius grows linearly\footnote{It is not clear what this entails physically. As explained by ACV~\cite{Amati:1987uf,Amati:1987wq,Amati:1988tn}, the stringy effects start to contribute when string length is comparable with Schwarzschild radius. At high energies, the stringy corrections only begin to contribute when Schwarzschild radius diverges, and this could potentially ruin the asymptotically flatness required for a scattering problem.} with $\sqrt{s}$. Hence at fixed angle, which is relevant for the Mellin transform, the scattering is not really absorbed/suppressed by the diffractive effect. However, the inelastic production of heavy strings/many strings does yields an interesting new suppression factor of the form 
\be\label{ineldamp}
\exp(-g_s\,\omega^2/\log\omega)\,,
\ee
which we have not encountered before.

We briefly comment on what this implies for the Mellin transform. Firstly, as explained around \eqref{EGampole1}, we can simply look at the Taylor expansion of $\exp(\omega^a/\log(\omega))$ near $\omega\sim 0$ and consider integrals like $\int_0^L \omega^{\Delta-1}\left( \frac{\omega^a}{\log(\omega)}\right)^n \d\omega$. Interestingly, this integral can be performed explicitly:
\begin{equation}
    \int_0^{L} \frac{\omega^{a+n-1}}{(\log(\omega))^n}\,   \d\omega=(-1)^n\, (a+n)^{n-1}\,\Gamma(1-n,-(a+n)\,\log L)\,,
\end{equation}
where $\Gamma(y,z):=\int_y^\infty x^{z-1}\,\e^{-x}\d x$ is the incomplete Gamma function. Although for $n=1$ a Gamma function $\Gamma(-a-1)$ appears, in general the pole structure is not a straightforward as the cases we encountered before. 

Another potential strategy would be to exploit the asymptotic series representation \eqref{2pt_function}, considering how the inelastic damping factor acts on the basis function $\cF(\beta,z)$ of \eqref{eq:calF4}, which is more likely an integral in $\beta$: $\partial_{\beta}^{-1}\Gamma(\beta/2+1)$. This will have the opposite effect as the exponentiated derivatives, shifting the poles to the right instead of left along the real axis. We would like to study this further in future works.


\subsection{Future directions}

We believe that there are many interesting future directions for exploration building upon the results of this paper. Here, we briefly touch upon a few of them. Of course, from the perspective of the celestial holography program, one of the most obvious questions is whether there is a 2d CFT which reproduces celestial probe or eikonal-probe amplitudes as 4-point or 2-point correlation functions, respectively. While portions of some perturbative celestial amplitudes have been recovered from dynamical CFTs, including certain chiral algebras and Liouville theory, it is not immediately clear how these constructions could produce the non-perturbative amplitudes obtained here. For eikonal-probe amplitudes, it would be interesting to explore the extent to which these can be recovered from marginal deformations of chiral CFTs, along the lines of the constructions in~\cite{Bu:2023vjt,Bu:2024cql} for perturbations away from self-duality. 

As we have emphasized, the pole structure in $\beta$ or $\Delta+\Delta'$ of our celestial amplitudes is subject to quantum corrections, as these poles arise from the small $\omega$ region of the Mellin transform which is outside the classical realm of the eikonal approximation. One could perhaps explore these quantum corrections explicitly by considering the inclusion of the first quantum correction to the eikonal approximation as encoded by $\delta_1$ of \eqref{eq:ACV}. 

In any case, real poles in real on-shell observables signals the existence of special on-shell states. For instance, when amplitudes solve the Bethe-Salpeter equation, the residues of these poles encodes information about the (modulus-squared) of bound state wavefunctions~\cite{Salpeter:1951sz,Gell-Mann:1951ooy,Nieuwenhuis:1993gh,Hoyer:2014gna,Hoyer:2016aew}. However, the poles we have observed in the conformal dimensions correspond to \emph{imaginary} energy poles. It was argued in~\cite{Paulos:2016but,Dorey:1996gd} that such poles indicate existence of unstable on-shell states which dissipate quickly over time. It would be of great interest to further study these states, especially in the black hole backgrounds.

Finally, we remark that it may be possible to make further progress by studying special cases of eikonal-probe scattering which are particularly simple, namely those corresponding to the \emph{self-dual sector}. There are many hints that self-duality is intricately connected with celestial holography. For instance, the self-dual Taub-NUT metric is the natural self-dual version of a Schwarzschild black hole~\cite{Hawking:1976jb,Crawley:2023brz}, and it admits an ultraboost which is a self-dual version of the Aichelberg-Sexl shockwave~\cite{Argurio:2008nb}. Recently, it was shown that probe graviton scattering (at arbitrary impact parameter) on the self-dual black hole takes a remarkably simple form in momentum space~\cite{Adamo:2023fbj}. Remarkably, the large distance geometry of the self-dual black hole extends to the near horizon region~\cite{Guevara:2023wlr}, meaning that the eikonal-probe amplitude is in fact the full probe amplitude. Analysing the Mellin transform of the simple probe formulae of~\cite{Adamo:2023fbj} could lead to further explicit analytic celestial amplitudes.


\paragraph{Acknowledgements}

The authors would like to thank Eduardo Casali for collaboration at an early stage of this project and some insightful comments on the draft, and Carlo Heissenberg for interesting discussions and insightful comments on the draft. PT would like to thank Giulia Isabella and Sasha Zhiboedov for discussions related to gravitational scattering and partial wave unitarity in gravity before this project was started. BZ would like to thank Stephan Stieberger and Tom Taylor for useful discussions. TA is supported by a Royal Society University Research Fellowship, the Leverhulme Trust grant RPG-2020-386, the Simons Collaboration on Celestial Holography MPS-CH-00001550-11 and the STFC consolidated grant ST/X000494/1. WB is supported by a Royal Society PhD Studentship. The work of PT has received funding from Agence Nationale de la Recherche (ANR), project ANR-22-CE31-0017. BZ is supported by the Royal Society.


\newpage
\appendix

\section{Analytic continuation of the eikonal Gamma function}\label{appendix:eik_gamma}

We define the eikonal Gamma function as the following integral:
\begin{equation}
   \Gamma_E(\beta,\lambda):= \int_0^\infty x^{\beta-1}\, \e^{-x}\,x^{-\lambda x}\,\d x\,.
\end{equation}
It is defined for $\beta>0,\lambda\geq0$, and reduces to the ordinary Gamma function for $\lambda=0$. For positive $\lambda$, $\Gamma_E$ can be continued in the whole complex $\beta$ plane apart from negative integers where we expect poles. 

The analytic continuation is done just like for the integral representation of the Gamma function. An integration by parts yields the alternative form
\begin{equation}
    \Gamma_E(\beta,\lambda) = -\frac{1}{\beta}\left((1+\lambda)\Gamma_E(\beta+1,\lambda)-\lambda \frac{\partial}{\partial \beta}\Gamma_E(\beta+1,\lambda) \right)\,.
    \label{eq:GammaE-AnCont}
\end{equation}
Hence, knowing $\Gamma_E$ for $0\leq \beta\leq 1$ allows is to be defined for $-1\leq \beta\leq0$, and so forth, on the whole complex plane. It is, in theory, possible to use this relation to measure the residues at negative integers, but we will proceed in a slightly easier way.

For negative integer $\beta=-n<0$, there are poles of order $n+1$. This is easy to see from a local perspective by Taylor expanding $x^{-\lambda x}$ near $x=0$. This produces an infinite sum $\sum_{n=0}^\infty \frac{(-\lambda x \log x)^n}{n!}$, and performing the formal manipulation of exchanging sum and integral, one obtains a sum of integrals whose integrands goes like $x^{\beta-1+n} (\log x)^n (-\lambda)^n/n!$ near $x=0$.
Using the simple integral
\begin{equation}
    \int_0^1 x^{n+\beta-1}\,(\log x)^n\, \d x = \frac{(-1)^n n!}{ (n+\beta)^{n+1}}\,,\quad n\in \mathbb{N}
\end{equation}
we see that, modulo a possible issue with exchanging summation and integration, $\Gamma_E$ should have a pole of order $n+1$ at $\beta=-n$, with residue should be $(-\lambda)^n/n!$.

\begin{figure}
    \centering
    \includegraphics{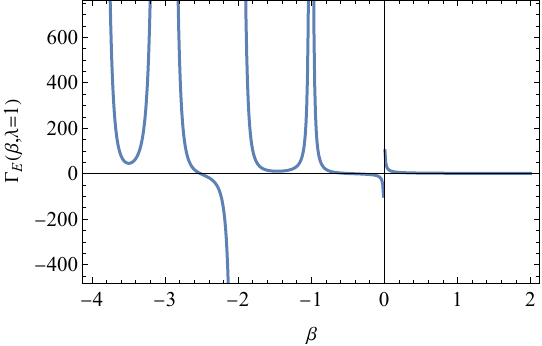}
    \caption{Analytically continued eikonal Gamma function for negative $\beta$.}
    \label{fig:plot-GammaE}
\end{figure}

This can be confirmed by implementing \eqref{eq:GammaE-AnCont} numerically, which we did. In Figure~\ref{fig:plot-GammaE} we depict $\Gamma_E$ at $\lambda=1$ as a function of $\beta$. This allowed us to check very precisely the conjecture above and verify that the residues of each pole are given by $\lambda^n$.

\renewcommand{\i}{{\rm i\,}}

Now we show that this integral can be continued so that 
\begin{equation}
    \int_0^\infty x^{\beta-1}\,\e^{\i x}\,x^{\i x}\,\d x\,,
\end{equation}
makes sense. It can be defined by a standard Cauchy contour. Reasoning backwards, take the integrand above and integrate it on a Cauchy contour as in fig.~\ref{fig:contour-eiko-gamma}. On the vertical axis, $x\to \i\,x$ and we recover an integral very close to the one we started with, so that
\begin{equation}
    \int_0^\infty x^{\beta-1}\, \e^{\i x}\,x^{\i x}\,\d x=\i^\beta \int_0^\infty x^{\beta-1}\, \e^{ -x(1+\i\frac{\pi}{2})}\,x^{- x}\,\d x\,,
\end{equation}
which is also obviously well defined. As a matter of fact, this integral is also easier to compute numerically and hence provides a way to cross-check the saddle point estimates we perform in the text in various places.

\subsection{Large $\beta$ limit} \label{app:d1}
We now look at the large $\beta$ limit of the eikonal Gamma function, slightly modified to be related to the complex one above:
\begin{equation}
   \Gamma_E(\beta,\lambda,\nu)= \int_0^\infty x^{\beta-1}\, \e^{-\nu  x}\,x^{-\lambda x}\, \d x
\end{equation}
where $\nu$ is an order 1 complex number. At large $\beta$, the power-law growth of $x^\beta$ compensates the exponential suppression of $\e^{-\nu x}\,x^{-\lambda x}$ and the integral is dominated by a saddle point $x_{*}$ at large $x$:
\begin{equation}
    \frac{\beta}{x_*}-\nu-\lambda\left(1+\log x_*\right)=0\,,
\end{equation}
where we have taken $\beta-1\to\beta$ in the $\beta\gg1$ limit. The various constants can be absorbed in the logarithm so that we solve in the end:
\begin{equation}
\label{eq:Delta-saddle}
    \frac{\beta}{\lambda} = x_*\, \log(c\, x_*)\,,\quad c:=\e^{1+\nu/\lambda}\,.
\end{equation}
This is solved by the product-log, or Lambert $W$, function, which is the solution to
\begin{equation}
    y e^y = v\,\to y = W_0(v)\,, \quad v>0\,,
\end{equation}
so that, calling $y=\log(w)$, one has
\begin{equation}
    w \log(w) =v \,\to \log w = W_0(v)\,.
\end{equation}
Since $w=\frac{v}{\log w }$ by definition, we find
\begin{equation}
    w=\frac{v}{W_0(v)}\,,
\end{equation}
so that \eqref{eq:Delta-saddle} is finally solved by
\begin{equation}\label{eq:productlog}
    x_*=\frac{\beta}{\lambda\, W_0(\tfrac{c\beta}{\lambda})}\,.
\end{equation}
Note that it were not for the $x^{-\lambda x}$ factor, the saddle would be located at $x=\beta$. We comment in passing that the Lambert $W$ function is not so often encountered in theoretical physics, with a notable exception being the QCD running coupling~\cite{Gardi:1998qr}.

Before proceeding, let us mention a few basic properties of the product-log function. By solving perturbatively \eqref{eq:productlog} at large $y$, we firstly have that
\begin{equation}
    y+\log(y) =\log(z)\implies y\simeq \log(z)-\log(\log(z))+\dots
\end{equation}
Therefore, the product-log is essentially a logarithm, up to $\log(\log)$ corrections\footnote{More precise bounds can be found for instance in~\cite{Hoorfar2008}.}.
Ignoring theses $\log(\log x)$ corrections, the saddle is located at
\begin{equation}
    x_*\sim\frac{\beta}{\lambda\, {\log\beta}}\,,
\end{equation}
up to subleading terms in $\beta/(\log\beta)^2$ and $ \beta\log(\log\beta)/(\log\beta)^2$.

\begin{figure}
    \centering
    \includegraphics{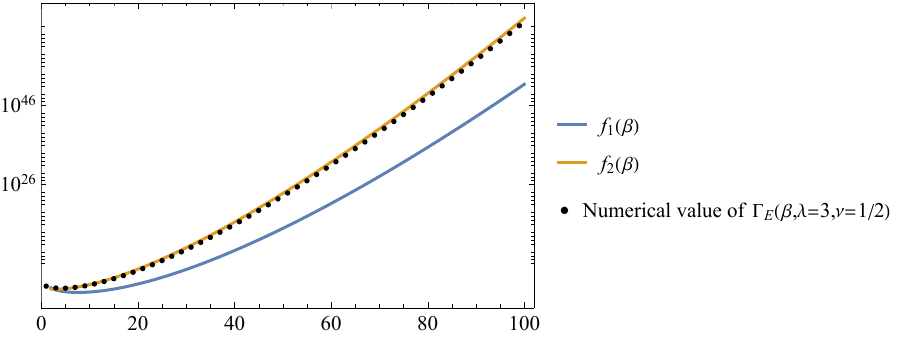}
    \caption{Plot of the numerical integration versus the two saddles of eq.\eqref{eq:f1} and \eqref{eq:f2}, corresponding respectively to the functions $f_1(\beta)$ and $f_2(\beta)$. Note that it is hard to see very clean convergence because in both $f_1$ and $f_2$ are neglected terms of order $1/(\log x)^2$ and $\log(\log x)/(\log x)^2$, which are very hard to detect at small values of $\beta\sim O(100)$, but the approximations are decent.}
    \label{fig:pl-saddle}
\end{figure}

On the saddle, the integrand then becomes:
\begin{equation}
    \exp\!\left( \beta \log x_* -\nu x_* - \lambda x_*\,\log x_* \right) 
    =\left(\frac{\beta}{\lambda\log(c\beta/\lambda)}\right)^\beta \,\e^{-\beta}\, \e^{\frac{\beta}{\log(c \beta/\lambda)}}\,. 
\end{equation}
Finally, the determinant contributes a factor of
\begin{equation}
    \left(\sqrt{-\frac{\beta}{x_*^2}+\frac{\lambda}{x_*}}\right)^{-1}\sim\frac{x_*}{\sqrt{\beta}}=
    \frac{\sqrt{\beta}}{\lambda\log(c\beta/\lambda)}\,.
\end{equation}
Overall we obtain that
\begin{equation}
\label{eq:f1}
    \Gamma_E(\beta,\lambda,\nu)\underset{\beta\to\infty}{\sim}
     \frac1\lambda \left(\frac{\beta}{\lambda\,\log(c\beta/\lambda)}\right)^\beta\, \e^{-\beta}\, \e^{\frac{\beta}{\log(c \beta/\lambda)}}\, \frac{\sqrt{\beta}}{\log(c\beta/\lambda)}\,.
\end{equation}
At this point, a slightly better approximation can be obtained by replacing the $1/\log(\cdot)$ by $1/W_0(\cdot)$:
\begin{equation}
\label{eq:f2}
    \Gamma_E(\beta,\lambda,\nu)\underset{\beta\to\infty}{\sim}
    \frac1\lambda \left(\frac{\beta}{\lambda\, W_0(c\beta/\lambda)}\right)^\beta\, \e^{-\beta}\, \e^{\frac{\beta}{ W_0(c \beta/\lambda)}}\, \frac{\sqrt{\beta}}{ W_0(c\beta/\lambda)}\,.
\end{equation}
For concreteness, we depict the numerical values of these approximations compared to the full numerical integral in fig.~\ref{fig:pl-saddle}.

\subsection{Asymptotic series representation}
In this section, we comment on the (non-)convergence of term-by-term integration suggested by the differential operator representation. That is, suppose one wishes to evaluate the eikonal Gamma function by expanding $x^{-\lambda x}$, and trading it for a differential operator:
\begin{equation}\label{ASRep1}
   \Gamma_E(\beta,\lambda)= \int_0^\infty x^{\beta-1}\, \e^{-x}\sum_{n=0}^\infty \frac{(-\lambda x\, \log x)^n}{n!}\,  \d x\overset{?}{=}
   \sum_{n=0}^\infty 
   \int_0^\infty x^{\beta-1}\, \frac{(-\lambda x\, (\log x))^n}{n!}\, \e^{-x}\, \d x\,.
\end{equation}
What we will give evidence that the series on the right-hand-side is asymptotic.

\begin{figure}
    \centering
    \includegraphics{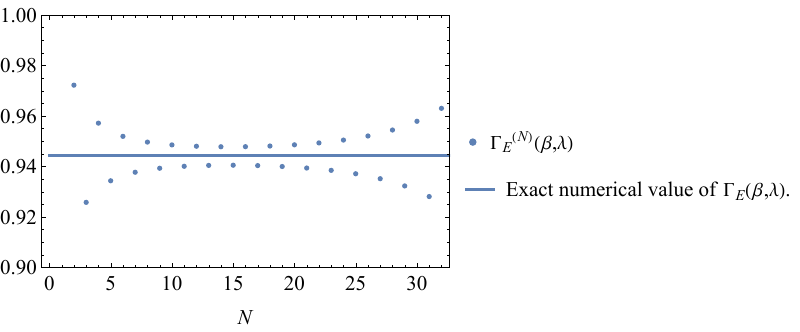}
    \caption{Convergence of the partial series summed until $N$ ($x$-axis) for $\beta=1$, $\lambda=1/4$. The thin line is the exact value of $\Gamma_E(1,1/4)$ computed numerically, the dots represent the partial series $\Gamma_E^{(N)} = \sum_{n=0}^N   \frac{(-\lambda)^n}{n!}\frac{\partial^n}{\partial \beta^n} \Gamma(\beta+n)$. At larger values of $\lambda$, the intermediate range convergence worsens.}
    \label{fig:asympt}
\end{figure}

Firstly, we can do a numerical experiment, and evaluate individually each term on the right-hand-side of \eqref{ASRep1}. These can be simply written as derivatives of a shifted Gamma function:
\begin{equation}\label{ASRep2}
    \int_0^\infty x^{\beta-1} \frac{(-\lambda x \,(\log x)^n}{n!}\, \e^{-x}\, \d x = 
    \frac{(-\lambda)^n}{n!}\,\frac{\partial^n}{\partial \beta^n} \Gamma(\beta+n)\,,
\end{equation}
and we can see numerically that this series is asymptotic. It begins by oscillating towards the true value, before eventually diverging. A typical situation is shown in fig.~\ref{fig:asympt}. 

This teaches us a lesson about differential operators. Remember that in the text, we proposed a representation of Mellin integrals of the form:
\begin{equation}
    \int_0^\infty \d\omega\,\omega^{\beta-1}\, \e^{-\omega}\, \e^{f(\omega,\log\omega)}\,,
\end{equation}
as essentially
\begin{equation}
   \e^{f(\partial_\beta,\e^{\partial_\beta})}\, \Gamma(\beta)\,,
\end{equation}
for some smooth function $f$. Here the exponentiated differential operator was to be understood as a Taylor expansion, producing ordinary $\beta$-derivatives and $\beta$-shifting operators $\e^{\partial_\beta}$. Clearly, this corresponds to the situation of \eqref{ASRep2} which we just described. The lesson is that one should be extremely cautious with this representation, which seems to give rise to an asymptotic series.

Empirically, it seems that whenever $\e^{f(\omega,\log\omega)}$ decays faster than $\e^{-\omega}$ (the other factor in the integrand of \eqref{ASRep2}, which is not Taylor expanded), the differential operator representation yields an asymptotic series. Clearly, we lose uniform convergence of the resulting Taylor series over the whole half-real line, but this is not enough to imply that integrating term-by-term will yield a divergent sum. We shall be content with the empirical observation for now.

However, our empirical investigation indicates that despite being asymptotic, this representation clearly knows about the correct result, as evidenced by its oscillations around the true numerical value. This could mean that it is amenable to resummation techniques, and it would be interesting to investigate this point further.

\newpage

\bibliographystyle{jhep}
\bibliography{c/cope}

\providecommand{\href}[2]{#2}\begingroup\raggedright\begin{thebibliography}{100}

\bibitem{Pasterski:2016qvg}
S.~Pasterski, S.-H. Shao, and A.~Strominger, {\it {Flat Space Amplitudes and Conformal Symmetry of the Celestial Sphere}},  {\em Phys. Rev. D} {\bf 96} (2017), no.~6 065026, [\href{http://arxiv.org/abs/1701.00049}{{\tt arXiv:1701.00049}}].

\bibitem{Pasterski:2017kqt}
S.~Pasterski and S.-H. Shao, {\it {Conformal basis for flat space amplitudes}},  {\em Phys. Rev. D} {\bf 96} (2017), no.~6 065022, [\href{http://arxiv.org/abs/1705.01027}{{\tt arXiv:1705.01027}}].

\bibitem{Pasterski:2017ylz}
S.~Pasterski, S.-H. Shao, and A.~Strominger, {\it {Gluon Amplitudes as 2d Conformal Correlators}},  {\em Phys. Rev. D} {\bf 96} (2017), no.~8 085006, [\href{http://arxiv.org/abs/1706.03917}{{\tt arXiv:1706.03917}}].

\bibitem{Arkani-Hamed:2020gyp}
N.~Arkani-Hamed, M.~Pate, A.-M. Raclariu, and A.~Strominger, {\it {Celestial amplitudes from UV to IR}},  {\em JHEP} {\bf 08} (2021) 062, [\href{http://arxiv.org/abs/2012.04208}{{\tt arXiv:2012.04208}}].

\bibitem{Pano:2024eek}
Y.~Pano and M.~Borji, {\it {Distributional Celestial Amplitudes}},  \href{http://arxiv.org/abs/2401.08877}{{\tt arXiv:2401.08877}}.

\bibitem{Stieberger:2018edy}
S.~Stieberger and T.~R. Taylor, {\it {Strings on Celestial Sphere}},  {\em Nucl. Phys. B} {\bf 935} (2018) 388--411, [\href{http://arxiv.org/abs/1806.05688}{{\tt arXiv:1806.05688}}].

\bibitem{Chang:2021wvv}
C.-M. Chang, Y.-t. Huang, Z.-X. Huang, and W.~Li, {\it {Bulk locality from the celestial amplitude}},  {\em SciPost Phys.} {\bf 12} (2022), no.~5 176, [\href{http://arxiv.org/abs/2106.11948}{{\tt arXiv:2106.11948}}].

\bibitem{Donnay:2023kvm}
L.~Donnay, G.~Giribet, H.~Gonz\'alez, A.~Puhm, and F.~Rojas, {\it {Celestial open strings at one-loop}},  {\em JHEP} {\bf 10} (2023) 047, [\href{http://arxiv.org/abs/2307.03551}{{\tt arXiv:2307.03551}}].

\bibitem{Castiblanco:2024hnq}
L.~Castiblanco, G.~Giribet, G.~Marin, and F.~Rojas, {\it {Celestial strings: field theory, conformally soft limits, and mapping the worldsheet onto the celestial sphere}},  \href{http://arxiv.org/abs/2405.01643}{{\tt arXiv:2405.01643}}.

\bibitem{Levy:1969cr}
M.~Levy and J.~Sucher, {\it {Eikonal approximation in quantum field theory}},  {\em Phys. Rev.} {\bf 186} (1969) 1656--1670.

\bibitem{DiVecchia:2023frv}
P.~Di~Vecchia, C.~Heissenberg, R.~Russo, and G.~Veneziano, {\it {The gravitational eikonal: from particle, string and brane collisions to black-hole encounters}},  \href{http://arxiv.org/abs/2306.16488}{{\tt arXiv:2306.16488}}.

\bibitem{Donnay:2018neh}
L.~Donnay, A.~Puhm, and A.~Strominger, {\it {Conformally Soft Photons and Gravitons}},  {\em JHEP} {\bf 01} (2019) 184, [\href{http://arxiv.org/abs/1810.05219}{{\tt arXiv:1810.05219}}].

\bibitem{Fan:2019emx}
W.~Fan, A.~Fotopoulos, and T.~R. Taylor, {\it {Soft Limits of Yang-Mills Amplitudes and Conformal Correlators}},  {\em JHEP} {\bf 05} (2019) 121, [\href{http://arxiv.org/abs/1903.01676}{{\tt arXiv:1903.01676}}].

\bibitem{Pate:2019mfs}
M.~Pate, A.-M. Raclariu, and A.~Strominger, {\it {Conformally Soft Theorem in Gauge Theory}},  {\em Phys. Rev. D} {\bf 100} (2019), no.~8 085017, [\href{http://arxiv.org/abs/1904.10831}{{\tt arXiv:1904.10831}}].

\bibitem{Adamo:2019ipt}
T.~Adamo, L.~Mason, and A.~Sharma, {\it {Celestial amplitudes and conformal soft theorems}},  {\em Class. Quant. Grav.} {\bf 36} (2019), no.~20 205018, [\href{http://arxiv.org/abs/1905.09224}{{\tt arXiv:1905.09224}}].

\bibitem{Puhm:2019zbl}
A.~Puhm, {\it {Conformally Soft Theorem in Gravity}},  {\em JHEP} {\bf 09} (2020) 130, [\href{http://arxiv.org/abs/1905.09799}{{\tt arXiv:1905.09799}}].

\bibitem{Guevara:2019ypd}
A.~Guevara, {\it {Notes on Conformal Soft Theorems and Recursion Relations in Gravity}},  \href{http://arxiv.org/abs/1906.07810}{{\tt arXiv:1906.07810}}.

\bibitem{Pate:2019lpp}
M.~Pate, A.-M. Raclariu, A.~Strominger, and E.~Y. Yuan, {\it {Celestial operator products of gluons and gravitons}},  {\em Rev. Math. Phys.} {\bf 33} (2021), no.~09 2140003, [\href{http://arxiv.org/abs/1910.07424}{{\tt arXiv:1910.07424}}].

\bibitem{Banerjee:2020kaa}
S.~Banerjee, S.~Ghosh, and R.~Gonzo, {\it {BMS symmetry of celestial OPE}},  {\em JHEP} {\bf 04} (2020) 130, [\href{http://arxiv.org/abs/2002.00975}{{\tt arXiv:2002.00975}}].

\bibitem{Fotopoulos:2019vac}
A.~Fotopoulos, S.~Stieberger, T.~R. Taylor, and B.~Zhu, {\it {Extended BMS Algebra of Celestial CFT}},  {\em JHEP} {\bf 03} (2020) 130, [\href{http://arxiv.org/abs/1912.10973}{{\tt arXiv:1912.10973}}].

\bibitem{Fotopoulos:2020bqj}
A.~Fotopoulos, S.~Stieberger, T.~R. Taylor, and B.~Zhu, {\it {Extended Super BMS Algebra of Celestial CFT}},  {\em JHEP} {\bf 09} (2020) 198, [\href{http://arxiv.org/abs/2007.03785}{{\tt arXiv:2007.03785}}].

\bibitem{Banerjee:2020zlg}
S.~Banerjee, S.~Ghosh, and P.~Paul, {\it {MHV graviton scattering amplitudes and current algebra on the celestial sphere}},  {\em JHEP} {\bf 02} (2021) 176, [\href{http://arxiv.org/abs/2008.04330}{{\tt arXiv:2008.04330}}].

\bibitem{Banerjee:2020vnt}
S.~Banerjee and S.~Ghosh, {\it {MHV gluon scattering amplitudes from celestial current algebras}},  {\em JHEP} {\bf 10} (2021) 111, [\href{http://arxiv.org/abs/2011.00017}{{\tt arXiv:2011.00017}}].

\bibitem{Adamo:2021zpw}
T.~Adamo, W.~Bu, E.~Casali, and A.~Sharma, {\it {Celestial operator products from the worldsheet}},  {\em JHEP} {\bf 06} (2022) 052, [\href{http://arxiv.org/abs/2111.02279}{{\tt arXiv:2111.02279}}].

\bibitem{Guevara:2021abz}
A.~Guevara, E.~Himwich, M.~Pate, and A.~Strominger, {\it {Holographic symmetry algebras for gauge theory and gravity}},  {\em JHEP} {\bf 11} (2021) 152, [\href{http://arxiv.org/abs/2103.03961}{{\tt arXiv:2103.03961}}].

\bibitem{Strominger:2021mtt}
A.~Strominger, {\it {$w_{1+\infty}$ Algebra and the Celestial Sphere: Infinite Towers of Soft Graviton, Photon, and Gluon Symmetries}},  {\em Phys. Rev. Lett.} {\bf 127} (2021), no.~22 221601, [\href{http://arxiv.org/abs/2105.14346}{{\tt arXiv:2105.14346}}].

\bibitem{Himwich:2021dau}
E.~Himwich, M.~Pate, and K.~Singh, {\it {Celestial operator product expansions and w$_{1+\infty}$ symmetry for all spins}},  {\em JHEP} {\bf 01} (2022) 080, [\href{http://arxiv.org/abs/2108.07763}{{\tt arXiv:2108.07763}}].

\bibitem{Schreiber:2017jsr}
A.~Schreiber, A.~Volovich, and M.~Zlotnikov, {\it {Tree-level gluon amplitudes on the celestial sphere}},  {\em Phys. Lett. B} {\bf 781} (2018) 349--357, [\href{http://arxiv.org/abs/1711.08435}{{\tt arXiv:1711.08435}}].

\bibitem{Fan:2022vbz}
W.~Fan, A.~Fotopoulos, S.~Stieberger, T.~R. Taylor, and B.~Zhu, {\it {Elements of celestial conformal field theory}},  {\em JHEP} {\bf 08} (2022) 213, [\href{http://arxiv.org/abs/2202.08288}{{\tt arXiv:2202.08288}}].

\bibitem{Casali:2022fro}
E.~Casali, W.~Melton, and A.~Strominger, {\it {Celestial amplitudes as AdS-Witten diagrams}},  {\em JHEP} {\bf 11} (2022) 140, [\href{http://arxiv.org/abs/2204.10249}{{\tt arXiv:2204.10249}}].

\bibitem{Fan:2022kpp}
W.~Fan, A.~Fotopoulos, S.~Stieberger, T.~R. Taylor, and B.~Zhu, {\it {Celestial Yang-Mills amplitudes and D = 4 conformal blocks}},  {\em JHEP} {\bf 09} (2022) 182, [\href{http://arxiv.org/abs/2206.08979}{{\tt arXiv:2206.08979}}].

\bibitem{deGioia:2022fcn}
L.~P. de~Gioia and A.-M. Raclariu, {\it {Eikonal approximation in celestial CFT}},  {\em JHEP} {\bf 03} (2023) 030, [\href{http://arxiv.org/abs/2206.10547}{{\tt arXiv:2206.10547}}].

\bibitem{Gonzo:2022tjm}
R.~Gonzo, T.~McLoughlin, and A.~Puhm, {\it {Celestial holography on Kerr-Schild backgrounds}},  {\em JHEP} {\bf 10} (2022) 073, [\href{http://arxiv.org/abs/2207.13719}{{\tt arXiv:2207.13719}}].

\bibitem{Banerjee:2023rni}
S.~Banerjee, R.~Mandal, A.~Manu, and P.~Paul, {\it {MHV Gluon Scattering in the Massive Scalar Background and Celestial OPE}},  \href{http://arxiv.org/abs/2302.10245}{{\tt arXiv:2302.10245}}.

\bibitem{Ball:2023ukj}
A.~Ball, S.~De, A.~Yelleshpur~Srikant, and A.~Volovich, {\it {Scalar-graviton amplitudes and celestial holography}},  {\em JHEP} {\bf 02} (2024) 097, [\href{http://arxiv.org/abs/2310.00520}{{\tt arXiv:2310.00520}}].

\bibitem{Crawley:2023brz}
E.~Crawley, A.~Guevara, E.~Himwich, and A.~Strominger, {\it {Self-dual black holes in celestial holography}},  {\em JHEP} {\bf 09} (2023) 109, [\href{http://arxiv.org/abs/2302.06661}{{\tt arXiv:2302.06661}}].

\bibitem{Stieberger:2022zyk}
S.~Stieberger, T.~R. Taylor, and B.~Zhu, {\it {Celestial Liouville theory for Yang-Mills amplitudes}},  {\em Phys. Lett. B} {\bf 836} (2023) 137588, [\href{http://arxiv.org/abs/2209.02724}{{\tt arXiv:2209.02724}}].

\bibitem{Taylor:2023bzj}
T.~R. Taylor and B.~Zhu, {\it {Celestial Supersymmetry}},  {\em JHEP} {\bf 06} (2023) 210, [\href{http://arxiv.org/abs/2302.12830}{{\tt arXiv:2302.12830}}].

\bibitem{Stieberger:2023fju}
S.~Stieberger, T.~R. Taylor, and B.~Zhu, {\it {Yang-Mills as a Liouville theory}},  {\em Phys. Lett. B} {\bf 846} (2023) 138229, [\href{http://arxiv.org/abs/2308.09741}{{\tt arXiv:2308.09741}}].

\bibitem{Giribet:2024vnk}
G.~Giribet, {\it {Remarks on celestial amplitudes and Liouville theory}},  \href{http://arxiv.org/abs/2403.03374}{{\tt arXiv:2403.03374}}.

\bibitem{tHooft:1987vrq}
G.~'t~Hooft, {\it {Graviton Dominance in Ultrahigh-Energy Scattering}},  {\em Phys. Lett. B} {\bf 198} (1987) 61--63.

\bibitem{Amati:1987uf}
D.~Amati, M.~Ciafaloni, and G.~Veneziano, {\it {Classical and Quantum Gravity Effects from Planckian Energy Superstring Collisions}},  {\em Int. J. Mod. Phys. A} {\bf 3} (1988) 1615--1661.

\bibitem{Kabat:1992tb}
D.~N. Kabat and M.~Ortiz, {\it {Eikonal quantum gravity and Planckian scattering}},  {\em Nucl. Phys. B} {\bf 388} (1992) 570--592, [\href{http://arxiv.org/abs/hep-th/9203082}{{\tt hep-th/9203082}}].

\bibitem{Adamo:2021rfq}
T.~Adamo, A.~Cristofoli, and P.~Tourkine, {\it {Eikonal amplitudes from curved backgrounds}},  {\em SciPost Phys.} {\bf 13} (2022), no.~2 032, [\href{http://arxiv.org/abs/2112.09113}{{\tt arXiv:2112.09113}}].

\bibitem{Duary:2022onm}
S.~Duary, {\it {Celestial amplitude for 2d theory}},  {\em JHEP} {\bf 12} (2022) 060, [\href{http://arxiv.org/abs/2209.02776}{{\tt arXiv:2209.02776}}].

\bibitem{Kapec:2022xjw}
D.~Kapec and A.~Tropper, {\it {Integrable field theories and their CCFT duals}},  {\em JHEP} {\bf 02} (2023) 128, [\href{http://arxiv.org/abs/2210.16861}{{\tt arXiv:2210.16861}}].

\bibitem{Bagchi:2022emh}
A.~Bagchi, S.~Banerjee, R.~Basu, and S.~Dutta, {\it {Scattering Amplitudes: Celestial and Carrollian}},  {\em Phys. Rev. Lett.} {\bf 128} (2022), no.~24 241601, [\href{http://arxiv.org/abs/2202.08438}{{\tt arXiv:2202.08438}}].

\bibitem{Donnay:2022aba}
L.~Donnay, A.~Fiorucci, Y.~Herfray, and R.~Ruzziconi, {\it {Carrollian Perspective on Celestial Holography}},  {\em Phys. Rev. Lett.} {\bf 129} (2022), no.~7 071602, [\href{http://arxiv.org/abs/2202.04702}{{\tt arXiv:2202.04702}}].

\bibitem{Donnay:2022wvx}
L.~Donnay, A.~Fiorucci, Y.~Herfray, and R.~Ruzziconi, {\it {Bridging Carrollian and celestial holography}},  {\em Phys. Rev. D} {\bf 107} (2023), no.~12 126027, [\href{http://arxiv.org/abs/2212.12553}{{\tt arXiv:2212.12553}}].

\bibitem{Mason:2023mti}
L.~Mason, R.~Ruzziconi, and A.~Yelleshpur~Srikant, {\it {Carrollian Amplitudes and Celestial Symmetries}},  \href{http://arxiv.org/abs/2312.10138}{{\tt arXiv:2312.10138}}.

\bibitem{Salzer:2023jqv}
J.~Salzer, {\it {An embedding space approach to Carrollian CFT correlators for flat space holography}},  {\em JHEP} {\bf 10} (2023) 084, [\href{http://arxiv.org/abs/2304.08292}{{\tt arXiv:2304.08292}}].

\bibitem{Nguyen:2023miw}
K.~Nguyen, {\it {Carrollian conformal correlators and massless scattering amplitudes}},  {\em JHEP} {\bf 01} (2024) 076, [\href{http://arxiv.org/abs/2311.09869}{{\tt arXiv:2311.09869}}].

\bibitem{Bagchi:2023cen}
A.~Bagchi, P.~Dhivakar, and S.~Dutta, {\it {Holography in Flat Spacetimes: the case for Carroll}},  \href{http://arxiv.org/abs/2311.11246}{{\tt arXiv:2311.11246}}.

\bibitem{Liu:2024nfc}
W.-B. Liu, J.~Long, and X.-Q. Ye, {\it {Feynman rules and loop structure of Carrollian amplitude}},  \href{http://arxiv.org/abs/2402.04120}{{\tt arXiv:2402.04120}}.

\bibitem{Have:2024dff}
E.~Have, K.~Nguyen, S.~Prohazka, and J.~Salzer, {\it {Massive carrollian fields at timelike infinity}},  \href{http://arxiv.org/abs/2402.05190}{{\tt arXiv:2402.05190}}.

\bibitem{Stieberger:2024shv}
S.~Stieberger, T.~R. Taylor, and B.~Zhu, {\it {Carrollian Amplitudes from Strings}},  \href{http://arxiv.org/abs/2402.14062}{{\tt arXiv:2402.14062}}.

\bibitem{Amati:1987wq}
D.~Amati, M.~Ciafaloni, and G.~Veneziano, {\it {Superstring Collisions at Planckian Energies}},  {\em Phys. Lett. B} {\bf 197} (1987) 81.

\bibitem{Amati:1988tn}
D.~Amati, M.~Ciafaloni, and G.~Veneziano, {\it {Can Space-Time Be Probed Below the String Size?}},  {\em Phys. Lett. B} {\bf 216} (1989) 41--47.

\bibitem{Amati:1990xe}
D.~Amati, M.~Ciafaloni, and G.~Veneziano, {\it {Higher Order Gravitational Deflection and Soft Bremsstrahlung in Planckian Energy Superstring Collisions}},  {\em Nucl. Phys. B} {\bf 347} (1990) 550--580.

\bibitem{Pasterski:2020pdk}
S.~Pasterski and A.~Puhm, {\it {Shifting spin on the celestial sphere}},  {\em Phys. Rev. D} {\bf 104} (2021), no.~8 086020, [\href{http://arxiv.org/abs/2012.15694}{{\tt arXiv:2012.15694}}].

\bibitem{Fernandes:2023ibv}
K.~Fernandes, F.-L. Lin, and A.~Mitra, {\it {Celestial Eikonal Amplitudes in the Near-Horizon Region}},  \href{http://arxiv.org/abs/2310.03430}{{\tt arXiv:2310.03430}}.

\bibitem{Gonzalez:2020tpi}
H.~A. Gonz\'alez, A.~Puhm, and F.~Rojas, {\it {Loop corrections to celestial amplitudes}},  {\em Phys. Rev. D} {\bf 102} (2020), no.~12 126027, [\href{http://arxiv.org/abs/2009.07290}{{\tt arXiv:2009.07290}}].

\bibitem{Torgerson:1966zz}
R.~Torgerson, {\it {Field-Theoretic Formulation of the Optical Model at High Energies}},  {\em Phys. Rev.} {\bf 143} (1966) 1194--1215.

\bibitem{Cheng:1969eh}
H.~Cheng and T.~T. Wu, {\it {High-energy elastic scattering in quantum electrodynamics}},  {\em Phys. Rev. Lett.} {\bf 22} (1969) 666.

\bibitem{Abarbanel:1969ek}
H.~D.~I. Abarbanel and C.~Itzykson, {\it {Relativistic eikonal expansion}},  {\em Phys. Rev. Lett.} {\bf 23} (1969) 53.

\bibitem{Cheng:1969tje}
H.~Cheng and T.~T. Wu, {\it {Impact factor and exponentiation in high-energy scattering processes}},  {\em Phys. Rev.} {\bf 186} (1969) 1611--1618.

\bibitem{Wallace:1973iu}
S.~J. Wallace, {\it {Eikonal expansion}},  {\em Annals Phys.} {\bf 78} (1973) 190--257.

\bibitem{Wallace:1977ae}
S.~J. Wallace and J.~A. McNeil, {\it {Relativistic Eikonal Expansion}},  {\em Phys. Rev. D} {\bf 16} (1977) 3565.

\bibitem{Tiktopoulos:1971hi}
G.~Tiktopoulos and S.~B. Treiman, {\it {Relativistic eikonal approximation}},  {\em Phys. Rev. D} {\bf 3} (1971) 1037--1040.

\bibitem{Eichten:1971kd}
E.~Eichten and R.~Jackiw, {\it {Failure of the eikonal approximation for the vertex function in a boson field theory}},  {\em Phys. Rev. D} {\bf 4} (1971) 439--443.

\bibitem{Kabat:1992pz}
D.~N. Kabat, {\it {Validity of the Eikonal approximation}},  {\em Comments Nucl. Part. Phys.} {\bf 20} (1992), no.~6 325--335, [\href{http://arxiv.org/abs/hep-th/9204103}{{\tt hep-th/9204103}}].

\bibitem{Guevara:2018wpp}
A.~Guevara, A.~Ochirov, and J.~Vines, {\it {Scattering of Spinning Black Holes from Exponentiated Soft Factors}},  {\em JHEP} {\bf 09} (2019) 056, [\href{http://arxiv.org/abs/1812.06895}{{\tt arXiv:1812.06895}}].

\bibitem{Chung:2018kqs}
M.-Z. Chung, Y.-T. Huang, J.-W. Kim, and S.~Lee, {\it {The simplest massive S-matrix: from minimal coupling to Black Holes}},  {\em JHEP} {\bf 04} (2019) 156, [\href{http://arxiv.org/abs/1812.08752}{{\tt arXiv:1812.08752}}].

\bibitem{Guevara:2019fsj}
A.~Guevara, A.~Ochirov, and J.~Vines, {\it {Black-hole scattering with general spin directions from minimal-coupling amplitudes}},  {\em Phys. Rev. D} {\bf 100} (2019), no.~10 104024, [\href{http://arxiv.org/abs/1906.10071}{{\tt arXiv:1906.10071}}].

\bibitem{Arkani-Hamed:2019ymq}
N.~Arkani-Hamed, Y.-t. Huang, and D.~O'Connell, {\it {Kerr black holes as elementary particles}},  {\em JHEP} {\bf 01} (2020) 046, [\href{http://arxiv.org/abs/1906.10100}{{\tt arXiv:1906.10100}}].

\bibitem{Moynihan:2019bor}
N.~Moynihan, {\it {Kerr-Newman from Minimal Coupling}},  {\em JHEP} {\bf 01} (2020) 014, [\href{http://arxiv.org/abs/1909.05217}{{\tt arXiv:1909.05217}}].

\bibitem{Haddad:2021znf}
K.~Haddad, {\it {Exponentiation of the leading eikonal phase with spin}},  {\em Phys. Rev. D} {\bf 105} (2022), no.~2 026004, [\href{http://arxiv.org/abs/2109.04427}{{\tt arXiv:2109.04427}}].

\bibitem{Bianchi:2023lrg}
M.~Bianchi, C.~Gambino, and F.~Riccioni, {\it {A Rutherford-like formula for scattering off Kerr-Newman BHs and subleading corrections}},  {\em JHEP} {\bf 08} (2023) 188, [\href{http://arxiv.org/abs/2306.08969}{{\tt arXiv:2306.08969}}].

\bibitem{Gatica:2023iws}
J.~P. Gatica, {\it {The Eikonal Phase and Spinning Observables}},  \href{http://arxiv.org/abs/2312.04680}{{\tt arXiv:2312.04680}}.

\bibitem{Luna:2023uwd}
A.~Luna, N.~Moynihan, D.~O'Connell, and A.~Ross, {\it {Observables from the Spinning Eikonal}},  \href{http://arxiv.org/abs/2312.09960}{{\tt arXiv:2312.09960}}.

\bibitem{Muzinich:1987in}
I.~J. Muzinich and M.~Soldate, {\it {High-Energy Unitarity of Gravitation and Strings}},  {\em Phys. Rev. D} {\bf 37} (1988) 359.

\bibitem{Giddings:2007bw}
S.~B. Giddings, D.~J. Gross, and A.~Maharana, {\it {Gravitational effects in ultrahigh-energy string scattering}},  {\em Phys. Rev. D} {\bf 77} (2008) 046001, [\href{http://arxiv.org/abs/0705.1816}{{\tt arXiv:0705.1816}}].

\bibitem{Giddings:2009gj}
S.~B. Giddings and R.~A. Porto, {\it {The Gravitational S-matrix}},  {\em Phys. Rev. D} {\bf 81} (2010) 025002, [\href{http://arxiv.org/abs/0908.0004}{{\tt arXiv:0908.0004}}].

\bibitem{Chen:2024iuv}
H.~Chen, R.~Karlsson, and A.~Zhiboedov, {\it {Energy correlations and Planckian collisions}},  \href{http://arxiv.org/abs/2404.15056}{{\tt arXiv:2404.15056}}.

\bibitem{Huang:2016tag}
Y.-t. Huang, O.~Schlotterer, and C.~Wen, {\it {Universality in string interactions}},  {\em JHEP} {\bf 09} (2016) 155, [\href{http://arxiv.org/abs/1602.01674}{{\tt arXiv:1602.01674}}].

\bibitem{DHoker:2019blr}
E.~D'Hoker and M.~B. Green, {\it {Exploring transcendentality in superstring amplitudes}},  {\em JHEP} {\bf 07} (2019) 149, [\href{http://arxiv.org/abs/1906.01652}{{\tt arXiv:1906.01652}}].

\bibitem{Brown:2004ugm}
F.~C.~S. Brown, {\it {Polylogarithmes multiples uniformes en une variable}},  {\em Compt. Rend. Math.} {\bf 338} (2004), no.~7 527--532.

\bibitem{Brown:2013gia}
F.~Brown, {\it {Single-valued Motivic Periods and Multiple Zeta Values}},  {\em SIGMA} {\bf 2} (2014) e25, [\href{http://arxiv.org/abs/1309.5309}{{\tt arXiv:1309.5309}}].

\bibitem{Stieberger:2013wea}
S.~Stieberger, {\it {Closed superstring amplitudes, single-valued multiple zeta values and the Deligne associator}},  {\em J. Phys. A} {\bf 47} (2014) 155401, [\href{http://arxiv.org/abs/1310.3259}{{\tt arXiv:1310.3259}}].

\bibitem{Stieberger:2014hba}
S.~Stieberger and T.~R. Taylor, {\it {Closed String Amplitudes as Single-Valued Open String Amplitudes}},  {\em Nucl. Phys. B} {\bf 881} (2014) 269--287, [\href{http://arxiv.org/abs/1401.1218}{{\tt arXiv:1401.1218}}].

\bibitem{Schlotterer:2018zce}
O.~Schlotterer and O.~Schnetz, {\it {Closed strings as single-valued open strings: A genus-zero derivation}},  {\em J. Phys. A} {\bf 52} (2019), no.~4 045401, [\href{http://arxiv.org/abs/1808.00713}{{\tt arXiv:1808.00713}}].

\bibitem{Vanhove:2018elu}
P.~Vanhove and F.~Zerbini, {\it {Single-valued hyperlogarithms, correlation functions and closed string amplitudes}},  {\em Adv. Theor. Math. Phys.} {\bf 26} (2022) 455--530, [\href{http://arxiv.org/abs/1812.03018}{{\tt arXiv:1812.03018}}].

\bibitem{Brown:2019wna}
F.~Brown and C.~Dupont, {\it {Single-valued integration and superstring amplitudes in genus zero}},  {\em Commun. Math. Phys.} {\bf 382} (2021), no.~2 815--874, [\href{http://arxiv.org/abs/1910.01107}{{\tt arXiv:1910.01107}}].

\bibitem{Alday:2022xwz}
L.~F. Alday, T.~Hansen, and J.~A. Silva, {\it {AdS Virasoro-Shapiro from single-valued periods}},  {\em JHEP} {\bf 12} (2022) 010, [\href{http://arxiv.org/abs/2209.06223}{{\tt arXiv:2209.06223}}].

\bibitem{Alday:2023mvu}
L.~F. Alday and T.~Hansen, {\it {The AdS Virasoro-Shapiro amplitude}},  {\em JHEP} {\bf 10} (2023) 023, [\href{http://arxiv.org/abs/2306.12786}{{\tt arXiv:2306.12786}}].

\bibitem{Alday:2024ksp}
L.~F. Alday and T.~Hansen, {\it {Single-valuedness of the AdS Veneziano amplitude}},  \href{http://arxiv.org/abs/2404.16084}{{\tt arXiv:2404.16084}}.

\bibitem{Verlinde:1991iu}
H.~L. Verlinde and E.~P. Verlinde, {\it {Scattering at Planckian energies}},  {\em Nucl. Phys. B} {\bf 371} (1992) 246--268, [\href{http://arxiv.org/abs/hep-th/9110017}{{\tt hep-th/9110017}}].

\bibitem{Cerulus:1964cjb}
F.~A. Cerulus and A.~Martin, {\it {A lower bound for large-angle elastic scattering at high energies}},  {\em Phys. Lett.} {\bf 8} (1964) 80--82.

\bibitem{Tourkine:2023xtu}
P.~Tourkine and A.~Zhiboedov, {\it {Scattering amplitudes from dispersive iterations of unitarity}},  {\em JHEP} {\bf 11} (2023) 005, [\href{http://arxiv.org/abs/2303.08839}{{\tt arXiv:2303.08839}}].

\bibitem{Buoninfante:2023dyd}
L.~Buoninfante, J.~Tokuda, and M.~Yamaguchi, {\it {New lower bounds on scattering amplitudes: non-locality constraints}},  {\em JHEP} {\bf 01} (2024) 082, [\href{http://arxiv.org/abs/2305.16422}{{\tt arXiv:2305.16422}}].

\bibitem{Haring:2023zwu}
K.~H\"aring and A.~Zhiboedov, {\it {The Stringy S-matrix Bootstrap: Maximal Spin and Superpolynomial Softness}},  \href{http://arxiv.org/abs/2311.13631}{{\tt arXiv:2311.13631}}.

\bibitem{Chowdhury:2019kaq}
S.~D. Chowdhury, A.~Gadde, T.~Gopalka, I.~Halder, L.~Janagal, and S.~Minwalla, {\it {Classifying and constraining local four photon and four graviton S-matrices}},  {\em JHEP} {\bf 02} (2020) 114, [\href{http://arxiv.org/abs/1910.14392}{{\tt arXiv:1910.14392}}].

\bibitem{Chandorkar:2021viw}
D.~Chandorkar, S.~D. Chowdhury, S.~Kundu, and S.~Minwalla, {\it {Bounds on Regge growth of flat space scattering from bounds on chaos}},  {\em JHEP} {\bf 05} (2021) 143, [\href{http://arxiv.org/abs/2102.03122}{{\tt arXiv:2102.03122}}].

\bibitem{Haring:2022cyf}
K.~H\"aring and A.~Zhiboedov, {\it {Gravitational Regge bounds}},  {\em SciPost Phys.} {\bf 16} (2024), no.~1 034, [\href{http://arxiv.org/abs/2202.08280}{{\tt arXiv:2202.08280}}].

\bibitem{Aichelburg:1970dh}
P.~C. Aichelburg and R.~U. Sexl, {\it {On the Gravitational field of a massless particle}},  {\em Gen. Rel. Grav.} {\bf 2} (1971) 303--312.

\bibitem{Dray:1984ha}
T.~Dray and G.~'t~Hooft, {\it {The Gravitational Shock Wave of a Massless Particle}},  {\em Nucl. Phys. B} {\bf 253} (1985) 173--188.

\bibitem{Jackiw:1991ck}
R.~Jackiw, D.~N. Kabat, and M.~Ortiz, {\it {Electromagnetic fields of a massless particle and the eikonal}},  {\em Phys. Lett. B} {\bf 277} (1992) 148--152, [\href{http://arxiv.org/abs/hep-th/9112020}{{\tt hep-th/9112020}}].

\bibitem{Adamo:2022rob}
T.~Adamo, A.~Cristofoli, and P.~Tourkine, {\it {The ultrarelativistic limit of Kerr}},  {\em JHEP} {\bf 02} (2023) 107, [\href{http://arxiv.org/abs/2209.05730}{{\tt arXiv:2209.05730}}].

\bibitem{Cornalba:2006xk}
L.~Cornalba, M.~S. Costa, J.~Penedones, and R.~Schiappa, {\it {Eikonal Approximation in AdS/CFT: From Shock Waves to Four-Point Functions}},  {\em JHEP} {\bf 08} (2007) 019, [\href{http://arxiv.org/abs/hep-th/0611122}{{\tt hep-th/0611122}}].

\bibitem{Cornalba:2006xm}
L.~Cornalba, M.~S. Costa, J.~Penedones, and R.~Schiappa, {\it {Eikonal Approximation in AdS/CFT: Conformal Partial Waves and Finite N Four-Point Functions}},  {\em Nucl. Phys. B} {\bf 767} (2007) 327--351, [\href{http://arxiv.org/abs/hep-th/0611123}{{\tt hep-th/0611123}}].

\bibitem{Cornalba:2007zb}
L.~Cornalba, M.~S. Costa, and J.~Penedones, {\it {Eikonal approximation in AdS/CFT: Resumming the gravitational loop expansion}},  {\em JHEP} {\bf 09} (2007) 037, [\href{http://arxiv.org/abs/0707.0120}{{\tt arXiv:0707.0120}}].

\bibitem{Bautista:2021wfy}
Y.~F. Bautista, A.~Guevara, C.~Kavanagh, and J.~Vines, {\it {Scattering in black hole backgrounds and higher-spin amplitudes. Part I}},  {\em JHEP} {\bf 03} (2023) 136, [\href{http://arxiv.org/abs/2107.10179}{{\tt arXiv:2107.10179}}].

\bibitem{Adamo:2023cfp}
T.~Adamo, A.~Cristofoli, A.~Ilderton, and S.~Klisch, {\it {Scattering amplitudes for self-force}},  {\em Class. Quant. Grav.} {\bf 41} (2024), no.~6 065006, [\href{http://arxiv.org/abs/2307.00431}{{\tt arXiv:2307.00431}}].

\bibitem{Vines:2017hyw}
J.~Vines, {\it {Scattering of two spinning black holes in post-Minkowskian gravity, to all orders in spin, and effective-one-body mappings}},  {\em Class. Quant. Grav.} {\bf 35} (2018), no.~8 084002, [\href{http://arxiv.org/abs/1709.06016}{{\tt arXiv:1709.06016}}].

\bibitem{Israel:1970kp}
W.~Israel, {\it {Source of the Kerr metric}},  {\em Phys. Rev. D} {\bf 2} (1970) 641--646.

\bibitem{Israel:1976vc}
W.~Israel, {\it {Line sources in general relativity}},  {\em Phys. Rev. D} {\bf 15} (1977) 935--941.

\bibitem{Balasin:1993kf}
H.~Balasin and H.~Nachbagauer, {\it {Distributional energy momentum tensor of the Kerr-Newman space-time family}},  {\em Class. Quant. Grav.} {\bf 11} (1994) 1453--1462, [\href{http://arxiv.org/abs/gr-qc/9312028}{{\tt gr-qc/9312028}}].

\bibitem{Garcia-Sepulveda:2022lga}
D.~Garc\'\i{}a-Sep\'ulveda, A.~Guevara, J.~Kulp, and J.~Wu, {\it {Notes on resonances and unitarity from celestial amplitudes}},  {\em JHEP} {\bf 09} (2022) 245, [\href{http://arxiv.org/abs/2205.14633}{{\tt arXiv:2205.14633}}].

\bibitem{Ghosh:2022net}
S.~Ghosh, P.~Raman, and A.~Sinha, {\it {Celestial insights into the S-matrix bootstrap}},  {\em JHEP} {\bf 08} (2022) 216, [\href{http://arxiv.org/abs/2204.07617}{{\tt arXiv:2204.07617}}].

\bibitem{Penrose:1980yx}
R.~Penrose, {\it {Null hypersurface initial data for classical fields of arbitrary spin and for general relativity}},  {\em Gen. Rel. Grav.} {\bf 12} (1980) 225--264.

\bibitem{Banerjee:2018gce}
S.~Banerjee, {\it {Null Infinity and Unitary Representation of The Poincare Group}},  {\em JHEP} {\bf 01} (2019) 205, [\href{http://arxiv.org/abs/1801.10171}{{\tt arXiv:1801.10171}}].

\bibitem{Banerjee:2019prz}
S.~Banerjee, S.~Ghosh, P.~Pandey, and A.~P. Saha, {\it {Modified celestial amplitude in Einstein gravity}},  {\em JHEP} {\bf 03} (2020) 125, [\href{http://arxiv.org/abs/1909.03075}{{\tt arXiv:1909.03075}}].

\bibitem{KoemansCollado:2019ggb}
A.~Koemans~Collado, P.~Di~Vecchia, and R.~Russo, {\it {Revisiting the second post-Minkowskian eikonal and the dynamics of binary black holes}},  {\em Phys. Rev. D} {\bf 100} (2019), no.~6 066028, [\href{http://arxiv.org/abs/1904.02667}{{\tt arXiv:1904.02667}}].

\bibitem{Adamo:2022ooq}
T.~Adamo and R.~Gonzo, {\it {Bethe-Salpeter equation for classical gravitational bound states}},  {\em JHEP} {\bf 05} (2023) 088, [\href{http://arxiv.org/abs/2212.13269}{{\tt arXiv:2212.13269}}].

\bibitem{DiVecchia:2022nna}
P.~Di~Vecchia, C.~Heissenberg, R.~Russo, and G.~Veneziano, {\it {The eikonal operator at arbitrary velocities I: the soft-radiation limit}},  {\em JHEP} {\bf 07} (2022) 039, [\href{http://arxiv.org/abs/2204.02378}{{\tt arXiv:2204.02378}}].

\bibitem{Bu:2023vjt}
W.~Bu and S.~Seet, {\it {A hidden 2d CFT for self-dual Yang-Mills on the celestial sphere}},  \href{http://arxiv.org/abs/2310.17457}{{\tt arXiv:2310.17457}}.

\bibitem{Bu:2024cql}
W.~Bu and S.~Seet, {\it {A systematic approach to celestial holography: a case study in Einstein gravity}},  \href{http://arxiv.org/abs/2404.04637}{{\tt arXiv:2404.04637}}.

\bibitem{Salpeter:1951sz}
E.~E. Salpeter and H.~A. Bethe, {\it {A Relativistic equation for bound state problems}},  {\em Phys. Rev.} {\bf 84} (1951) 1232--1242.

\bibitem{Gell-Mann:1951ooy}
M.~Gell-Mann and F.~Low, {\it {Bound states in quantum field theory}},  {\em Phys. Rev.} {\bf 84} (1951) 350--354.

\bibitem{Nieuwenhuis:1993gh}
T.~Nieuwenhuis, J.~A. Tjon, and Y.~A. Simonov, {\it {Relativistic two-body bound state calculations beyond the ladder approximation}},  \href{http://arxiv.org/abs/hep-ph/9309267}{{\tt hep-ph/9309267}}.

\bibitem{Hoyer:2014gna}
P.~Hoyer, {\it {Bound states -- from QED to QCD}},  2, 2014.
\newblock \href{http://arxiv.org/abs/1402.5005}{{\tt arXiv:1402.5005}}.

\bibitem{Hoyer:2016aew}
P.~Hoyer, {\it {Lectures on Bound states}},  5, 2016.
\newblock \href{http://arxiv.org/abs/1605.01532}{{\tt arXiv:1605.01532}}.

\bibitem{Paulos:2016but}
M.~F. Paulos, J.~Penedones, J.~Toledo, B.~C. van Rees, and P.~Vieira, {\it {The S-matrix bootstrap II: two dimensional amplitudes}},  {\em JHEP} {\bf 11} (2017) 143, [\href{http://arxiv.org/abs/1607.06110}{{\tt arXiv:1607.06110}}].

\bibitem{Dorey:1996gd}
P.~Dorey, {\it {Exact S matrices}},  in {\em {Eotvos Summer School in Physics: Conformal Field Theories and Integrable Models}}, pp.~85--125, 8, 1996.
\newblock \href{http://arxiv.org/abs/hep-th/9810026}{{\tt hep-th/9810026}}.

\bibitem{Hawking:1976jb}
S.~W. Hawking, {\it {Gravitational Instantons}},  {\em Phys. Lett. A} {\bf 60} (1977) 81.

\bibitem{Argurio:2008nb}
R.~Argurio, F.~Dehouck, and L.~Houart, {\it {Boosting Taub-NUT to a BPS NUT-wave}},  {\em JHEP} {\bf 01} (2009) 045, [\href{http://arxiv.org/abs/0811.0538}{{\tt arXiv:0811.0538}}].

\bibitem{Adamo:2023fbj}
T.~Adamo, G.~Bogna, L.~Mason, and A.~Sharma, {\it {Scattering on self-dual Taub-NUT}},  {\em Class. Quant. Grav.} {\bf 41} (2024), no.~1 015030, [\href{http://arxiv.org/abs/2309.03834}{{\tt arXiv:2309.03834}}].

\bibitem{Guevara:2023wlr}
A.~Guevara and U.~Kol, {\it {Self Dual Black Holes as the Hydrogen Atom}},  \href{http://arxiv.org/abs/2311.07933}{{\tt arXiv:2311.07933}}.

\bibitem{Gardi:1998qr}
E.~Gardi, G.~Grunberg, and M.~Karliner, {\it {Can the QCD running coupling have a causal analyticity structure?}},  {\em JHEP} {\bf 07} (1998) 007, [\href{http://arxiv.org/abs/hep-ph/9806462}{{\tt hep-ph/9806462}}].

\bibitem{Hoorfar2008}
A.~Hoorfar and M.~Hassani, {\it {Inequalities on the Lambert W function and hyperpower function}},  {\em J. Ineq. in Pure and Appl. Math.} {\bf 9} (2008) 51.

\end{thebibliography}\endgroup

\end{document}